\documentclass[a4paper,11pt]{article}
\usepackage{aaskaiid}
\usepackage{xcolor}
\usepackage{enumitem}
\usepackage{wrapfig}
\usepackage{comment}

\newcommand{\ag}[1]{{\color{violet}[AG: #1]}}

\DeclareUnicodeCharacter{2212}{-}

\title{Synergies for the Epoch of Reionization and Cosmic Dawn}
\ShortTitle{Reionization and Cosmic Dawn Synergies}
\author[1]{Anirban Chakraborty}
\author[1]{Tirthankar R. Choudhury}
\author[2]{Kanan K. Datta}
\author[3,4,5]{Pratika Dayal}
\author[6,7]{Jiten Dhandha}
\author[8]{Samuel Gagnon-Hartman}
\author[9,10]{Sambit K. Giri}
\author[11]{Adélie Gorce}
\author[12]{Caroline Heneka}
\author[13]{Anne Hutter}
\author[14]{Barun Maity}
\author[15]{Suman Majumdar}
\author[16]{Andrei Mesinger}
\author[17, 18]{Kana Moriwaki}
\author[19]{Chandra Shekhar Murmu}
\author[20]{Yuxiang Qin}
\author[21]{Shintaro Yoshiura}
\author[22]{Erik Zackrisson}

\affiliation[1]{National Centre for Radio Astrophysics, Tata Institute of Fundamental Research, Pune 411007, India}
\affiliation[2]{Relativity \& Cosmology Research Centre, Department of Physics, Jadavpur University,
Kolkata 700032, India}
\affiliation[3]{Canadian Institute for Theoretical Astrophysics, 60 St George St, University of Toronto, Toronto, ON M5S 3H8, Canada}

\affiliation[4]{David A. Dunlap Department of Astronomy and Astrophysics, University of Toronto, 50 St George St, Toronto ON M5S 3H4, Canada}

\affiliation[5]{Department of Physics, 60 St George St, University of Toronto, Toronto, ON M5S 3H8, Canada}
\affiliation[6]{Institute of Astronomy, University of Cambridge, Madingley Road, Cambridge CB30HA, United Kingdom}

\affiliation[7]{Kavli Institute for Cosmology, Madingley Road, Cambridge CB30HA, United Kingdom}
\affiliation[8]{Scuola Normale Superiore di Pisa, Piazza dei Cavallieri 7, 56126 Pisa, Italy}
\affiliation[9]{Department of Astronomy and Oskar Klein Centre, AlbaNova, Stockholm University, SE-10691 Stockholm, Sweden}

\affiliation[10]{Van Swinderen Institute for Particle Physics and Gravity, University of Groningen, Nijenborgh 3, 9747 AG Groningen, The Netherlands}
\affiliation[11]{Université Paris-Saclay, CNRS, Institut d'Astrophysique Spatiale, 91405, Orsay, France}

\affiliation[12]{Institut f\"ur Theoretische Physik, Universit\"at Heidelberg, Philosophenweg 16, 69120 Heidelberg, Germany}
\affiliation[13]{Institute for Astronomy, University of Vienna, T\"urkenschanzstrasse 17, A-1180 Vienna, Austria}
\affiliation[14]{Max-Planck-Institut für Astronomie, Königstuhl 17, D-69117 Heidelberg, Germany}
\affiliation[15]{Department of Astronomy, Astrophysics \& Space Engineering,
Indian Institute of Technology Indore,
Indore 453552, India}
\affiliation[16]{Department of Physics and Astronomy ``Ettore Majorana", University of Catania, Via Santa Sofia 64, 95123  Catania, Italy}
\emailAdd{andrei.mesinger@dfa.unict.it}

\affiliation[17]{Research Center for the Early Universe, Graduate School of Science, The University of Tokyo, 7-3-1 Hongo, Bunkyo, Tokyo 113-0033, Japan}

\affiliation[18]{Department of Physics, Graduate School of Science, The University of Tokyo, 7-3-1 Hongo, Bunkyo, Tokyo 133-0033, Japan}
\affiliation[19]{Astrophysics Research Centre of the Open University (ARCO) \& Department of Natural Sciences,
The Open University of Israel,
1 University Road, POBox808, Ra’anana 4353701, Israel}
\affiliation[20]{Research School of Astronomy and Astrophysics, Australian National University, Canberra, ACT 2611, Australia}
\affiliation[21]{Mizusawa VLBI Observatory, National Astronomical Observatory of Japan, 2-21-1 Osawa, Mitaka, Tokyo 181-8588, Japan}
\affiliation[22]{Observational Astrophysics, Department of Physics and Astronomy, Uppsala University, Box 516, SE-751 20 Uppsala, Sweden}
\abstract{
Synergies with other instruments will be {\it essential} in making, verifying, and interpreting a detection of the cosmic 21-cm signal from the Epoch of Reionization (EoR) and Cosmic Dawn (CD) with the SKA-low telescope.
Such synergies can (i) provide prior information about galaxies and the intergalactic medium (IGM) during the EoR/CD; (ii) pave the road to a first 21cm detection by mitigating foregrounds and systematics through cross-correlations; and (iii) give complimentary physical insights into the galaxy -- IGM connection.
Here we review the current state of synergies and discuss what observations will best compliment SKA-low EoR/CD observations.
}

\begin{document}
\maketitle

\section{Introduction}

The Epoch of Reionization (EoR) remains one of the central frontiers of modern cosmology. After decades of research, we are beginning to constrain the timing of the main phase of reionization (e.g. \citealt{2016A&A...596A.108P, Greig2017MNRAS.466.4239G, mason2018universe, bosman2022hydrogen, qin2025percent}). Yet, we still lack a clear understanding of how to connect this cosmic milestone to the populations of stars and black holes that powered it.

The most powerful probe of this era is the 21-cm line from neutral hydrogen, which traces fluctuations in ionization, temperature, and density of the intergalactic medium (IGM). Current interferometers aim to detect the 21-cm power spectrum statistically (e.g. \citealt{Paciga11, Acharya2024MNRAS.534L..30A, Yoshiura2021A17, HERACollaboration2023}). SKA-Low in Australia promisses to deliver the first 3D tomographic map of the early Universe, providing unprecedented insights into the astrophysics of galaxies, the IGM, and cosmology itself (e.g. \citealt{Mesinger20}).

However, high–signal-to-noise (S/N) 21-cm maps of the EoR and the preceding Cosmic Dawn (CD) are still years away. Initial detections will likely involve only a few low-S/N power-spectrum modes. Given the novelty of the observation, confirming that such signals are truly cosmological will be challenging. The most robust way to validate them—and to enhance S/N—is through cross-correlation with independent tracers of known cosmic origin.

Cross-correlations not only provide a sanity check on early detections but also yield cleaner cosmological signals, since foregrounds and instrumental systematics in different datasets are typically uncorrelated. 
Moreover, multi-wavelength observations of the same volume provide complementary insights into the physics governing the galaxy -- IGM connection during the EoR.  
Eventually, with an expanded SKA (Phase 2), it will be possible to correlate full 3D images from different probes, enabling detailed studies of individual ionized or heated regions—comparing their 21-cm morphology with the brightest galaxies they host, as observed by {\it JWST} and future optical/IR facilities.

%The 21-cm signal depends on the IGM’s density, temperature, and ionized fraction. During the dark ages, it traces matter fluctuations directly; after the emergence of the first galaxies, radiative coupling and heating introduce non-linearities across scales. Since galaxies are biased tracers of the matter field, any large-scale tracer should in principle correlate with the 21-cm signal. The nature and strength of this correlation encode rich information about early galaxy populations.

Cross-correlating real datasets is technically challenging: the ``footprints” of different surveys must overlap in both angular and redshift space. In 21-cm interferometry, foregrounds combined with the instrument response contaminates a ``wedge” region in Fourier space (e.g. \citealt{LT11, Vedantham12, Morales12}).  The remaining so-called "EoR window" spans large transverse scales combined with small line of sight scales (i.e. low $k_\perp$ + high $k_{||}$). Hence, ideal cross-correlation partners are large-volume surveys with accurate redshift localization.  Possible candidates for cross-correlations include:
\begin{enumerate}
\item {\it Cosmic backgrounds} — Integrated backgrounds such as the CMB, NIR, and X-ray backgrounds include emission from $z > 5.5$ and should correlate with the EoR/CD 21-cm signal (e.g. \citealt{MaCiardi_2018, ZhouLaPlante_2025}). However, their broad redshift integration limits overlap with 21-cm modes lost to foreground removal.\\
\item {\it Resolved galaxies} — Galaxy maps are natural 21-cm counterparts, but surveys at $z > 5.5$ must balance wide fields, accurate redshifts, and sufficient source densities. Narrow-band dropout techniques, grism surveys, and slit spectroscopic follow-ups are the most promising approaches (e.g. \citealt{Vrbanec_2020, Heneka_2020, Hutter_2023b}).\\
\item {\it Line Intensity Mapping (LIM)} — Mapping of lines such as [C II], [O III], CO, and Ly$\alpha$ can trace the combined  emission from unresolved galaxies. LIM offers large volumes and good redshift precision, making it an ideal complement to 21-cm tomography (e.g. \citealt{LIM:2017}). However, LIM observations at high redshift remain untested, and expected line luminosities are faint (e.g. \citealt{Yue+2015})\\
\item Quasar spectra — The Lyman-$\alpha$ forest in quasar spectra can be cross-correlated with the 21-cm forest, providing complementary UV and radio absorption probes of the same IGM regions \citep{Bhagwat22}. This requires radio-loud quasars at high redshift and a sufficiently cold IGM for 21-cm absorption.
\end{enumerate}

In this chapter we review the current state of synergies, highlighting how they are shaping our knowledge of EoR galaxies and the IGM.  We also discuss future potentials for cross-correlation with the cosmic 21-cm signal.  We end with a summary of early science targets that could be used to make and verify the first 21-cm detection of the EoR with SKA.

%The final phase change of our Universe, the so-called Epoch of Reionization (EoR), remains at the forefront of modern cosmology.  After decades of studies, we are starting to narrow-in on the timing of the bulk of reionization (e.g. Qin et al. 2021, Bosman et al. 2022).  However, we still do not really know how to connect this final phase change of our Universe to the populations of stars and black holes that drive it.  

%In addition to providing an invaluable sanity check on preliminary claims of a 21-cm detection, cross-correlations can also improve the S/N. The cross power spectrum could provide a cleaner probe of the cosmological signal, since the foregrounds and systematics of different datasets are typically not correlated (e.g. Chime collaboration 2023).   Eventually, with a significant expansion of the SKA (referred to as SKA Phase 2) we should be able to correlate images (i.e. including phases) of different datasets.  This would allow us to study individual ionized or heated regions, comparing their tomography (obtained with SKA) to the brightest galaxies they contain (obtained with optical/IR telescopes like JWST).

%%%%%%%%%%%%%%%%%%%%%%%%%%%%%%%%%%%%%%%%%%%%%%%%%%%%%%%%%%%%%%%%%%%%%%%%%%%%%%%%%
%%%%%%%%%%%%%%%%%%%%%%%%%%%%%%%%%  WHAT DO WE KNOW?  %%%%%%%%%%%%%%%%%%%%%%%%%%%%%%%%%
\section{Informing our understanding}

In this section we discuss how multi-wavelength studies of the EoR are shaping our understanding of the dominant physical processes.

%%%%%%%%%%%%%%%%%%%%%%%%%%%%%%%%%  NATURE OF SOURCES %%%%%%%%%%%%%%%%%%%%%%%%%%%%%%%%%%%%
\subsection{The nature of sources}

The sources of reionization remain a key outstanding issue in the field (see reviews in, e.g., \citealt{Mesinger16, dayal_2018}). This is because the relative contributions of sources (including low versus high-mass star-forming galaxies and black holes) depend on a number of poorly understood parameters including {\it (i)}: the halo-mass dependence of star formation and black hole accretion; {\it (ii)}: intrinsic stellar populations (star formation history, initial mass function, metallicity) that determine the output of ionizing photons; {\it (iii)}: the line-of-sight and galaxy property dependent escape fraction of Lyman continuum photons into the IGM ($f_{\rm esc}$); and {\it (iv)}: the clustering of sources that determines both the reionization morphology and its feedback on the baryonic contents of low-mass halos. 

\begin{figure}[h]
    \centering
    \includegraphics[width=\columnwidth]{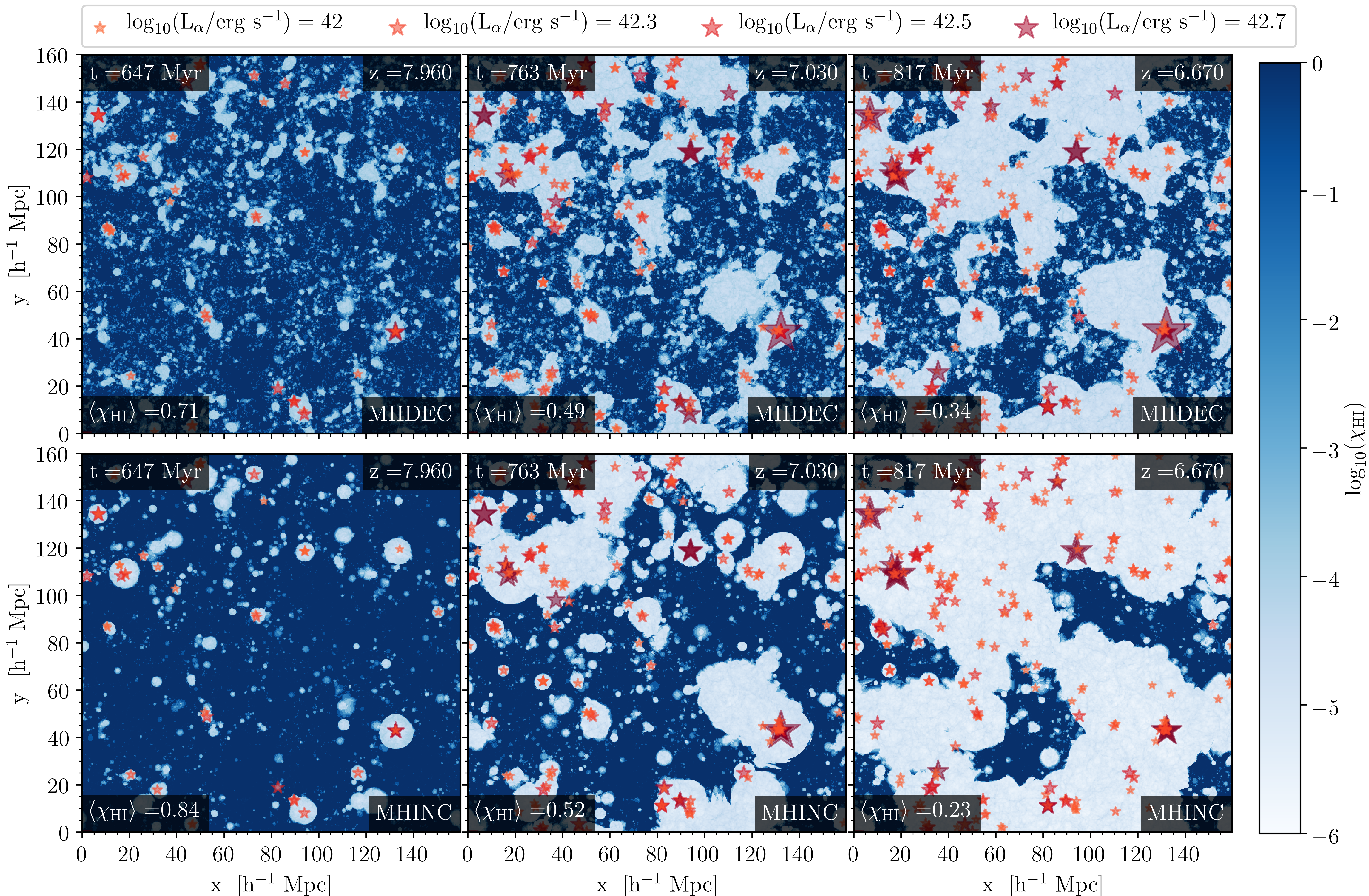}
    \caption{The progress of reionization and the associated evolution of Lyman Alpha Emitter (LAE) visibility between $z\sim 8$ and $z \sim 6.6$, from left to right, as marked \citep{hutter_2023}. The upper and lower rows show results for scenarios in which $f_{\rm esc}$ decreases ({\sc mhdec}) and increases ({\sc mhinc}) with an increase in halo mass, respectively. EoR morphology is much richer in small-scale structures when low-mass sources dominate the process ({\sc mhdec}) as opposed to being much more biased when rarer higher-mass sources drive the process ({\sc mhinc}) (e.g. \citealt{McQuinn07}).}
    \label{astraeus_fesc_lya}
\end{figure}

The James Webb Space Telescope (JWST) has been transformational in extending constraints on the star-forming galaxy population well within the first billion years between $z \sim 10-15$ \citep{adams_2024, harikane2024_spec,donnan_2024} with even some preliminary estimates of the ultra-violet luminosity function out to $z \sim 15-20$ \citep{kokorev_2025, whitler_2025,castellano_2025}, going down to systems with absolute UV magnitudes as faint as ${\rm M_{UV}} \sim$ -15 to -17 i.e. $0.005 L_*$ \citep{atek_2024}. We also have tentative hints on the stellar mass functions  \citep{weibel_2024, harvey_2025}, mass-metallicity relations \citep{curti_2024, chemerynska_2024} and UV spectral slopes \citep{austin_2024, rinaldi_2024}. Together, these build a picture of robustly star-forming, low-metallicity, young systems within the first billion years. Crucially, JWST observations have also yielded hints on the ionizing photon production efficiency ($\xi_{\rm ion}$ in units of Hz s$^{-1}$) which is the production rate of ionizing photons per unit UV luminosity: while theoretical estimates suggested $\log\xi_{\rm ion} \sim 25.2$, observations imply $\log \xi_{\rm ion}\sim25.8$ \citep{simmonds_2024, begley_2024, atek_2024}. Observational works also imply the $\log \xi_{\rm ion}$ scales inversely with the galaxy luminosity, decreasing from  25.6 to 25 as $M_{\rm UV}$ decreases from $\sim -18$ to $\sim -22$ at $z \sim 5$ \citep{llerena_2024}. %Such high values of $\xi_{ion}$ necessitate low values of $f_{esc}\sim {\rm few}\%$ to yield reionization histories in accord with both cosmological and galaxy observables \citep{dayal_2025, ferrara_2025,2024JCAP...07..078C}. 
Furthermore, the dependence of $f_{\rm esc}$ on galaxy properties (and hence redshift) is one of the most crucial parameters for reionization (e.g. \citealt{Maity_Choudhury2022a, qin2025percent}).  Together, these properties determine not just the EoR morphology but also the distribution of galaxies visible as Lyman Alpha Emitters (e.g. \citealt{sobacchi2015clustering, hutter_2023}; c.f. Fig. \ref{astraeus_fesc_lya}). 
Matching JWST UV luminosity functions at $z\ge10$, along with data related to reionization history, might point to enhanced star-formation efficiency and halo-mass–dependent escape fractions (e.g. \citealt{qin2025percent}). Depending on the galaxy model, these trends might be in tension with observed galaxy clustering \citep{2024JCAP...07..078C,2025arXiv250307590C}.

JWST observations have also revealed a large population of AGN in the early universe (z$>$4) which, although fainter, are much more abundant than luminous quasars found by pre-JWST surveys \citep{maiolino2024_jades,taylor2024}. Surprisingly, the luminosity function of AGN spectroscopically identified by JWST is about one or two orders of magnitude higher the extrapolation of the luminosity function of pre-JWST discovered quasars at $z \sim 5-7$.  AGN are estimated to make up a few percent of the galaxy population at the bright end \citep{greene2024, kokorev2024, akins2024,matthee2024}, contributing up to $18-30\%$ of the UV luminosity function \citep{scholtz2025}. Such high AGN number densities have prompted a reevaluation of their role in reionization.  Models that have AGN dominating the EoR must however make extreme assumptions such as high escape fractions, high occupancy, soft (stellar like) spectra and Compton thick disks (e.g. \citealt{madau2024, grazian2024}).  Most works still credit faint galaxies as dominant sources of the EoR \citep{robertson2015, madau2017,Qin2017MNRAS.472.2009Q, atek_2024, dayal_2025, qin2025percent}.% and in agreement with previous findings that AGN mostly contribute to the patchiness of reionization in its end stages \citep{chardin2017}, and do not have any appreciable effect on the 21cm power spectrum \citep{trebitsch2023}.

Finally, we expect that the heating of the IGM to temperatures exceeding the CMB likely preceded the bulk of the EoR.  This  was expected given the observed scaling relations between X-ray luminosities and star formation rates in local galaxies (e.g. \citealt{Fragos12, Kaur23}), and has recently been confirmed observationally using 21-cm upper limits (e.g. \citealt{HERACollaboration2023}).  This heating is most likely driven by X-rays emitted by metal-poor X-ray binary stars in the first galaxies  \citep{Kaur23, Sartorio23}.  However, more exotic sources of heating could be provided by accreting primordial black holes (e.g. \citealt{dayal2024_pbh, zhang2025}), decaying dark matter (e.g. \citealt{LS18, Facchinetti24}), annihilating dark matter (e.g. \citealt{EMF14, Honorez16}), or cosmic rays \citep{SS15, Leite17, JN18, Gessey-Jones23}.

%However, these galaxy datasets will need to be correlated with the underlying ionization fields, or anti-correlated with the 21cm field, to shed light on the key reionization sources. As discussed in Sections \ref{sec:3-2,sec:4-1} that follow, with the smallest smearing our of power in redshift space, an anti-correlation between line emitters (including Lyman Alpha Emitters, [CII] and OIII emitters) will be the most crucial probe of the key reionization sources. 

%%%%%%%%%%%%%%%%%%%%%%%%%%%%%%%%%  IGM  %%%%%%%%%%%%%%%%%%%%%%%%%%%%%%%%%%%%%%%
\subsection{The state of the IGM}
\label{sec:state_IGM}

% Discuss constraints on the ionization and thermal history of the IGM, ionized bubbles and neutral island topology, ultraviolet background, ... 

% Caroline: Direct inference of xHI, constrained modelling of xHI from density field with NNs; plus a note in this section on measuring the spin temperature + fluctuations (linking to cross-correlation in following sections)

\begin{figure*}
    \centering
    \includegraphics[width=1.0\linewidth]{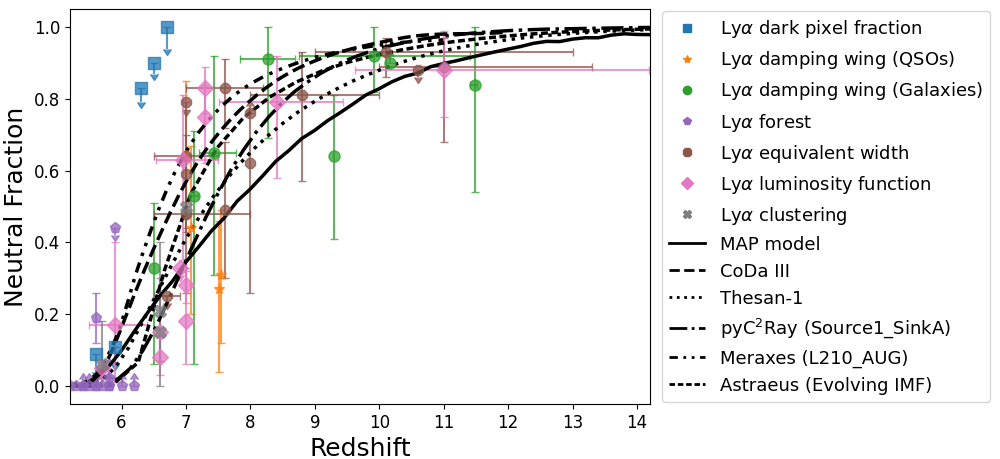}
    % \vspace{-0.6cm}  
    \caption{
    Constraints on the reionization history of the intergalactic medium from multiple observational probes. Different Lyman-$\alpha$ observations constrain the ionization fraction as a function of redshift, including: the dark pixel fraction in quasar spectra \citep{mcgreer2015model,jin2023nearly}, Lyman-$\alpha$ damping wing profiles \citep{greig2022igm,curtis2023spectroscopic,hsiao2024jwst,umeda2024jwst,mason2025constraints}, the transmitted flux in the Lyman-$\alpha$ forest \citep{yang2020measurements,bosman2022hydrogen,spina2024damping,zhu2024damping}, Lyman-$\alpha$ equivalent widths \citep{mason2018universe,mason2019inferences,jung2020texas,bolan2022inferring,bruton2023universe,nakane2024lyalpha,tang2024lyalpha,jones2025jades}, the luminosity function evolution of Lyman-$\alpha$ emitters \citep{inoue2018silverrush,morales2021evolution,wold2022lager,umeda2025silverrush,kageura2025census}, and their spatial clustering properties \citep{sobacchi2015clustering,ouchi2018systematic,umeda2025silverrush}. We show the maximum-a-posteriori (MAP) model from \citet{qin2025percent} that was inferred from Lyman alpha forest data, together with UV LFs and the CMB optical depth. State-of-the-art numerical simulations, including \texttt{CoDa III} \citep{lewis2022short}, \texttt{Thesan-1} \citep{garaldi2022thesan}, \texttt{pyC$^2$Ray} \citep[Source1\_SinkA;][]{giri202421}, \texttt{Meraxes} \citep[L210\_AUG;][]{balu2023thermal}, \texttt{Astraeus} \citep[Evolving IMF;][]{hutter2025astraeus}, are consistent with these observational constraints.
    }
    \label{fig:reionization_history_constraints}
\end{figure*}

%Prior to direct measurements of the EoR using the 21-cm line, recent observational studies have provided significant constraints on various properties of the IGM, including ionization and thermal histories and the topology of ionized and neutral regions, helping us prepare models and strategies for upcoming measurements with SKA-Low. %\ch{Suggestion added, rest of paragraph} 
%One of the early science outcomes expected from SKA-Low is the redshift evolution of the mean neutral fraction of the IGM, $x_\mathrm{HI}$.  %This can be evolution from power spectra \citep[e.g.,][]{pietschke2025direct}, as currently being tested with the SKA Science Data Challenge SDC3b Inference (Bonaldi+, in prep.). 
%In addition, cross-correlation with galaxies, prominently LAEs, is expected to facilitate a detection and both determine the progress of reionization via $x_\mathrm{HI}$ (see section~\ref{sec:21cm-galaxy} for references and discussion) and the thermal state of the IGM sensitive to the spin temperature $T_\mathrm{S}$ \citep{Heneka_2020}. %CH: for (potential) Ts-dependency of 21cm-LAE cross see e.g. Fig.3 in arXiv:2004.10097
%A core goal of SKA-Low is to constrain the progress of reionization and IGM properties directly from tomographic imaging \citep[e.g.,][]{giri2018optimal}, where machine learning is foreseen to play a crucial role in direct reconstruction and inference \citep{bianco2021deep,bianco2024deep,bianco2025deep}. %CH: can ref and cite what's mentioned in other chapters and sections, plus we'll have a paper on this out in about a week (Pietschke+)

Prior to direct measurements of the EoR using the 21-cm line, we must rely on alternative probes provided by the damping wing absorption in galaxy and quasar spectra, the Lyman alpha forest, and the CMB.  In Fig.~\ref{fig:reionization_history_constraints}, we show the latest constraints on the ionization history derived from several different types of measurements. These constraints are roughly consistent with each other, implying an early start to the EoR (e.g. \citealt{curtis2023spectroscopic,tang2024lyalpha}) and  a late end extending below $z\sim 6$ \citep[e.g.,][]{kulkarni2019large,bosman2022hydrogen}. The maximum-a-posteriori (MAP) model from \citet{qin2025percent}, inferred from the Lyman alpha forest, combined with UV LFs and the CMB optical depth, is also shown in the figure. While these constraints remain broad, they provide valuable guidance for model development. The figure shows several reionization models developed using different frameworks, including hydrodynamical simulations—\texttt{CoDa III} \citep{lewis2022short} and \texttt{Thesan-1} \citep{garaldi2022thesan}—and post-processing of dark matter N-body simulations—\texttt{pyC$^2$2Ray} \citep[Source1\_SinkA;][]{giri202421}, \texttt{Meraxes} \citep[L210\_AUG;][]{balu2023thermal}, and \texttt{Astraeus} \citep[Evolving IMF;][]{hutter2025astraeus}. However, most simulations must tune their ionizing escape fraction in order to get reasonable EoR histories, limiting their predictive powers.

The thermal history of the IGM has been extensively studied at $z \lesssim 6$ using the Lyman-$\alpha$ forest \citep{schaye2000thermal, theuns2002constraints, bolton2010first,bolton2012improved, boera2019revealing, walther2019new, gaikwad2020probing, Maity_Choudhury2022a,Maity_Choudhury2022b}. Complementary limits on the thermal state of the neutral IGM at $z>10$ have been derived from 21-cm observations \citet{HERACollaboration2023}, implying an additional source of heating beyond the well-known processes of adiabatic and Compton cooling/heating.

\subsection{Inference using complementary data}\label{subsec:2_3_joint}

\begin{figure*}
    \centering
    \includegraphics[width=.8\linewidth]{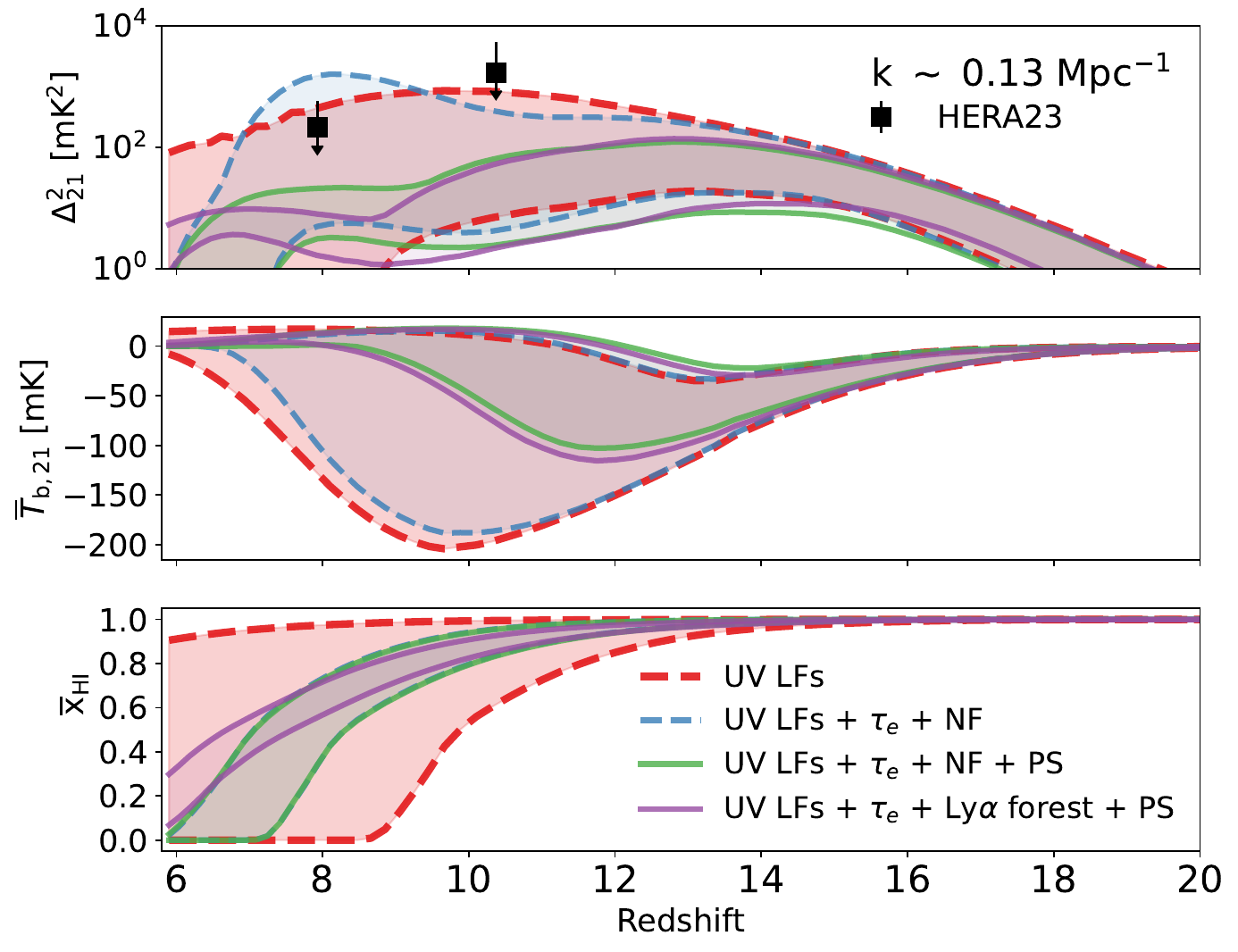}
    % \vspace{-0.6cm}  
    \caption{\label{fig:IGM_inference}Constraints on the redshift evolution of the 21-cm power spectrum at $k=0.13$ cMpc$^{-1}$, the 21-cm global signal and mean neutral fraction, top to bottom. The shaded regions and areas enclosed between paired curves indicate 95\% confidence intervals inferred from different combinations of observational probes, including the high-redshift galaxy UV luminosity functions (UF LFs), CMB optical depth ($\tau_e$), neutral/dark fraction (NF) or Lyman-$\alpha$ (Ly$\alpha$) forest statistics from high-redshift quasar spectra, and the latest 21-cm power spectrum upper limits from \citet[][PS]{HERACollaboration2023}. This figure is obtained using the {\tt 21cmFAST} simulation code \citep{Mesinger2011MNRAS.411..955M,Murray2020JOSS....5.2582M}, and adapted from \citet{Breitman2024MNRAS.527.9833B}.}
\end{figure*}

%this could be a separate section if it grows too large...
Complementary datasets are essential for constraining the state of the IGM during the EoR. They help in breaking degeneracies among key astrophysical parameters and provide independent insights into different phases and components of the reionization process. In recent years, there has been a surge in EoR-relevant observations across a wide range of probes (see some examples in Fig. \ref{fig:reionization_history_constraints}), including: 
\begin{enumerate}
\item primary and secondary anisotropies of the cosmic microwave background (CMB; \citealt{2016A&A...596A.108P,2021ApJ...908..199R})
\item the large-scale fluctuations in Lyman-alpha forest trasmission (\citealt{fan2006constraining,bosman2022hydrogen}) and its associated dark gaps \citep{Mesinger2010MNRAS.407.1328M,Zhu2021ApJ...923..223Z};
\item the damping-wing absorption in high-redshift quasar \citep{Greig2017MNRAS.466.4239G,Banados2018Natur.553..473B,Davies2018} and galaxy spectra \citep{curtis2023spectroscopic,umeda2024jwst};
\item the luminosity functions of broad band dropouts \citep{inoue2018silverrush,morales2021evolution}, and Lyman-alpha emitters \citep{sobacchi2015clustering,ouchi2018systematic};
\item upper limits on the 21-cm power spectrum from radio interferometers \citep{Trott2020DeepObservations,HERACollaboration2023,MertensMevius_2025,Nunhokee2025ApJ...989...57N}.%; (vi) A putative detection of the sky-averaged global 21-cm signal (\citealt{Bowman2018}), although with a lack of confirmation \citep{Singh2022NatAs...6..607S}.
\end{enumerate}

Theoretical frameworks have advanced in parallel, with different approaches ranging from analytic to semi-numerical to full radiative transfer simulations (see a couple of examples in Fig. \ref{fig:reionization_history_constraints}). These frameworks aim to capture inhomogeneous reionization driven by physically motivated galaxy formation models, and are increasingly capable of integrating multiple complementary probes in their inference pipelines.

What do these datasets imply for the evolution of the IGM and the corresponding 21cm signal?  In Figure \ref{fig:IGM_inference} we show four inference results, varying which datasets are included in the likelihood: (i) UV luminosity functions (UV LFs) of galaxies alone \citep{Bouwens2015ApJ...803...34B,Bouwens2016ApJ...830...67B,Oesch2018ApJ...855..105O}; (ii) CMB optical depth \citep{2016A&A...596A.108P,QinPoulin_2020} and Dark Fraction upper limits on the neutral hydrogen fraction \citep{mcgreer2015model}, in addition  to the UV LFs; (iii) 21-cm power spectrum upper limits from HERA \citep{HERACollaboration2023}, included in addition to the above; (iv) substituting the Dark Fraction with the latest measurements of the Lyman alpha effective optical depth \citep{bosman2022hydrogen,qin2025percent}.

The UV LFs alone rule out extreme scenarios in which the IGM is heated or ionized either too early or not at all.  Adding the CMB optical depth and Dark Fraction constrains the neutral fraction evolution, restricting the reionization mid-point to a narrower window of $7<z<8$. While this combination impacts the positive envelope of the mean 21cm signal (when it is in emission), and so is difficult to see in the figure in which the dynamic range of the mean signal is dominated by the absorption trough, we do see a large impact on the 21cm power spectrum.  This is especially notable  at $z<7$, by excluding models that reionize too early or too late.

In contrast, the inclusion of current 21-cm PS upper limits from HERA does not significantly alter the inferred ionization history, but substantially tightens the posterior on the 21cm brightness temperature. These limits effectively rule out "cold reionization" models \citep{MEWH14} in which the neutral IGM patches are significantly colder than the CMB (or other) radio background, and thus provide a strong brightness temperature contrast with the ionized IGM patches during the EoR, resulting in a high amplitude power spectrum. Finally, replacing the Dark Fraction with the most recent Lyman alpha effective optical depth measurements not only shifts the posterior distribution of the neutral fraction but also leads to noticeable changes in the resulting 21-cm PS, reflecting the updated understanding of the timing of reionization. With a peak in the 21cm power spectrum at later times, where instrumental noise is lower, there appears to be a growing opportunity to detect the 21-cm power spectrum in the near future.

Observations of present-day radiation backgrounds, including the X-ray background observed by Chandra and the radio background observed by ARCADE, can provide additional constraints on the galaxy population during the EoR and CD.  These are broadly consistent with the constraints above, but can additionally disfavor very early heating and strong radio excess (e.g. \citealt{Dhandha2025MNRAS.542.2292D,Dhandha2025MNRAS.544.1608D}), with the caveat that connecting the high-$z$ population to the $z=0$ background is sensitive to the assumed X-ray and radio spectral energy distributions (e.g. \citealt{Mirocha25, Katz25}).

%A similar analysis using the 21cmSPACE code (including diffuse X-ray limits from Chandra and radio limits from ARACADE2/LWA1, but not neutral fraction constraints) disfavour very early heating or radio excess scenarios, placing stringent but optimistic limits on the 21-cm global signal at $z\approx10-15$ and of the power spectrum at $z \approx 15$ \citep{Dhandha2025MNRAS.542.2292D,Dhandha2025MNRAS.tmp.1629D}.

%%%%%%%%%%%%%%%%%%%%%%%%%%%%%%%%%%%%%%%%%%%%%%%%%%%%%%%%%%%%%%%%%%%%%%%%%%%%%%%%%
%%%%%%%%%%%%%%%%%%%%%%%%%%%%%%%%%  CROSS  %%%%%%%%%%%%%%%%%%%%%%%%%%%%%%%%%%%%%%%
\section{Statistical cross-correlation}

As mentioned in the introduction, cross-correlating early SKA data with other observations will be important as a sanity check for first claims of a detection as well as helping mitigate unknown systematics.  Here we discuss possible targets for such cross-correlation studies, commenting on their detectability.

%%%%%%%%%%%%%%%%%%%%%%%%%%%%%%%%%  CROSS w CRB %%%%%%%%%%%%%%%%%%%%%%%%%%%%%%%%%%%%%%%
\subsection{Cosmic Background Radiation}
\label{sec:3_1}
\subsubsection{CMB}

We have seen in Sec.~\ref{subsec:2_3_joint} that measurements of the large-scale CMB polarisation and temperature power spectra can be combined with astrophysical and 21\,cm measurements to sharpen our picture of the reionisation history \citep{QinPoulin_2020}.
Indeed, fluctuations in the electron density during reionization lead to distinct scaling relations in both the patchy–kSZ and B‑mode signals -- relations that, when combined with 21\,cm observations, can robustly break degeneracies between optical depth, reionization duration, and bubble morphology \citep{2021MNRAS.500..232P}.
%The Epoch of Reionization imprints the CMB observables through two main physical phenomena. 
\paragraph{Thomson scattering.}

%In a bid to limit the impact of foregrounds and systematics plaguing high-redshift 21\,cm observations, a
Thomson scattering of CMB photons off electrons produced by reionisation dampen CMB temperature anisotropies and enhance its polarisation anisotropies. Through this effect, many CMB observables can be cross-correlated with 21\,cm observations to isolate the reionisation contribution in both signals.
\citet{TashiroAghanim_2008} have shown that the cross-correlation of CMB $E$-mode polarisation anisotropies with 21\,cm brightness fluctuations, on large scales, can accurately constrain the reionisation history: the peak of the cross-correlation spectrum is maximum when about half of the IGM is ionised, and there is a damping in the spectrum that depends on the duration of reionisation. However, the faintness of the signal makes it difficult to observe with experiments such as LOFAR and Planck \citep{TashiroAghanim_2010}. Only an ambitious survey carried out with the SKA-\textit{Low}, where several fields are observed for 1000~hours to cover 2\% of the sky could lead to a signal-to-noise ratio of one -- and this is ignoring residual foregrounds and systematics polluting the cross-signal.

The patchiness of the reionisation process leads to spatial fluctuations in this optical depth which can be leveraged to constrain reionisation morphology. We expect a lot of information about reionisation to be enclosed in the spatial fluctuations of $\tau$ \citep{DvorkinSmith_2009, GluscevicKamionkowski_2013, MeerburgMeyers_2017}. Their study has been gaining momentum in the past few years, namely as a tracer of post-reionisation large-scale structures \citep{SchuttManiyar_2024}. The detectability of these fluctuations has been established in the literature, either alone, with telescopes such as the now canceled CMB-Stage~4, or of even better sensitivity, such as CMB-HD \citep{RoyLapi_2018,RoyKulkarni_2021}; or sooner in cross-correlation with, e.g., gravitational lensing with SPT-3G \citep{BianchiniMillea_2023}. During reionisation, spatial fluctuations in $\tau$ and in the 21\,cm signal are expected to be anti-correlated, as scattering comes from ionised regions whilst 21\,cm brightness stems from neutral Hydrogen \citep{HolderIliev_2007}. Some works in the literature have shown the anti-correlation shape and amplitude to depend not only on the reionisation history, but also on the details of the physics of reionisation, such as ionised bubble sizes \citep{MeerburgDvorkin_2013, RoyLapi_2020}. Assuming that the measurement is limited by instrumental noise and cosmic variance rather than foregrounds and systematics, \citet{RoyLapi_2020} find a cumulative SNR of 11.5 when cross-correlating observations from the SKA and the Simons Observatory, already online, provided observations cover at least 10\% of the sky.
Additionally, the reionisation patchiness leads to the production of faint $B$-modes, which are directly related to $C_\ell^{\tau\tau}$, and, thus, trace reionisation history and morphology \citep{Mukherjee_2019,Choudhury_2021,RoyKulkarni_2021,Jain2023, Jain2024_BB}. Despite similar arguments as $\tau \times 21$\,cm correlations, so far, no forecast for the measurements of 21\,cm $\times$ EoR-induced $B$-modes angular cross-spectra has been performed.

\paragraph{The kinetic Sunyaev-Zel'dovich effect.} CMB photons can interact with electrons from reionisation that have a bulk velocity with respect to the CMB rest-frame through a Doppler shift process known as the kinetic Sunyaev-Zel'dovich (kSZ) effect. Additional temperature anisotropies are produced by this interaction, both on large scales \citep{AlvarezKomatsu_2006, AdsheadFurlanetto_2008} and at scales corresponding to the typical size of ionised bubbles during reionisation (multipoles $\ell \sim 3000$). Extensive work has been performed to forecast the potential of cross-correlating this signal with 21\,cm brightness temperature, both with numerical simulations \citep{SalvaterraCiardi_2005,JelicZaroubi_2010, MMS12, LaPlanteLidz_2020} or analytical models \citep{Cooray_2004,McQuinnFurlanetto_2005,TashiroAghanim_2011}. In every configuration, the cross-signal is found to be a good tracer of reionisation history and morphology, mainly through its shape and sign flips, as well as potentially of cosmic magnetism \citep{Kunze_2023}. On large scales, although \citet{TashiroAghanim_2010} have found the signal to be detectable with a cumulative SNR of about 10 (2) for a combination of Planck and SKA (LOFAR) data, a detection has not yet been achieved, namely because of strong foreground pollution in the 21\,cm data, as demonstrated for MWA observations in \citet{YoshiuraIchiki_2019}. 
Such observations on very large scales (several degrees) would be limited by the smaller field-of-view of the SKA. 
 On smaller scales, the anti-correlation traces the size of ionised bubbles as it is stronger for larger bubbles, and peaks at a multipole one can relate to the typical bubble size \citep{SalvaterraCiardi_2005,JelicZaroubi_2010,TashiroAghanim_2011}. However, the signal is overall faint as ionised regions are equally likely to be moving towards or away from the observer\footnote{This cancellation is partially avoided on large scales, where the signal is, however, fainter \citep{AdsheadFurlanetto_2008,MaHelgason_2018}.} \citep{AlvarezKomatsu_2006}. Different options have been discussed in the literature to limit this effect:
\begin{itemize}
    \item Cross-correlating kSZ and 21\,cm maps that have been integrated over the line of sight. However, since the homogeneous (patchy) kSZ anisotropies (anti-)correlate with the 21\,cm signal, the two contributions tend to cancel each other \citep{JelicZaroubi_2010}.
    \item Two-point correlations after squaring the kSZ signal -- making sure the primary CMB component and other sources of unwanted noise are filtered out before performing the squaring. In the absence of 21\,cm foregrounds, this signal is detectable at the $\gtrsim 35\sigma$ level for 10~hours of integration time with the SKA \citep{MaHelgason_2018}. \citet{ZhouLaPlante_2025} have more recently proposed to also square the 21\,cm signal to compensate for a filter applied to remove the brightest foregrounds contributions whilst retaining cosmological information. Such a signal will also be detectable with a combination of SKA and Stage~3 CMB experiments, such as SO or SPT-3G.
    \item Three-point correlations, two points being taken in the kSZ field\footnote{Note that this statistic is related to the previous one by a simple integral.} \citep{Cooray_2004}, whose intensity is related to the reionisation history. In the absence of foreground contamination, this signal is detectable with SKA and SO -- at more than $20\sigma$ for the cumulative signal \citep{LaPlanteLidz_2020}. However, the kSZ signal is sensitive only to Fourier modes with long-wavelength line-of-sight components, which are most strongly polluted by foregrounds in 21\,cm data sets. 
\end{itemize}
Overall, a wealth of information is enclosed in the kSZ$\times$21\,cm signal, which is highly detectable by the SKA in combination with both current and future CMB experiments, in a framework where uncertainties are dominated by noise. Further work must be carried out to assess the true detectability of the signal -- and reach detection, e.g., taking into account detailed foreground mitigation techniques, both in CMB and in 21\,cm data analysis. 

\begin{comment}
\ag{
\begin{itemize}[nolistsep]
    %\item Quick description of the different CMB observables of EoR/CD: optical depth (inc. fluctutions), $B$-modes, kinetic SZ
    %\item What cross-correlations can tell us about EoR? ($\tau\tau$x21, kSZx21, $BB$x21 \citep{KadotaOoba_2019}, order 2 or +)
    \item Which experiments overlap with SKA-Low and detectability
    \item What could we wish for in the future? (i.e. design your perfect CMB experiment for cross-correlations with SKA)
\end{itemize}}
\end{comment}

\subsubsection{The near infrared background}

%The sources of cosmic reionisation have not yet been fully identified. Among others, the contribution from faint galaxies is an active subject of research. Until recently, extrapolating the faint-end of the UV galaxy luminosity functions beyond the detection limit ($M_\mathrm{lim} = -15$, mostly with the HST) was necessary to reionise the Universe in time to match the low CMB optical depth measured by Planck \citep{BouwensIllingworth_2015, GorceDouspis_2018}. On the other end, recent observations with the JWST on small, extremely deep fields seem to indicate that ultra-faint high-redshift galaxies have a much higher ionising emissivity than anticipated \citep{atek_2024, simmonds_2024}, which could boost the reionisation process \citep{munoz_2024}. In order to understand if the JWST-observed fields are representative of the large-scale distribution of faint galaxies and if these results on the ionising emissivity of ultra-faint galaxies can be generalised, a solution lies in the near-infrared background (NIRB). 
The cumulative ionising (UV) light of faint galaxies at $z\sim10-20$ could form background radiation, redshifted to infrared frequencies. \citet{SunMirocha_2021} have shown that Pop~III
stars can leave characteristic spectral signatures on this near infrared background (NIRB) spectrum, depending on the conditions of their formation. Observed by SPHEREx, such features could provide constraints on the abundance and formation history of Pop~III stars at high redshifts. %thanks to their strong Ly$\alpha$ emission 

%The high-redshift 21\,cm signal stems from neutral regions that have not yet seen galaxy formation, whilst the IRB is sourced by faint galaxies. Therefore, the two 
The NIRB and the 21cm signal trace opposite phases of the IGM on large scales during reionisation (we expect them to be to be anti-correlated) and can be combined to provide powerful constraints on the reionisation history, as the NIRB alone cannot provide redshift information \citep{ShanQin_2009,Mao_2014,FernandezZaroubi_2014}. However, the cross-signal is difficult to access in observations as NIRB fluctuations probe large-scale modes which are dominated by foregrounds in 21\,cm data. Similarly to what has been proposed for CMB cross-correlations, the 21\,cm signal can be squared after filtering out foreground-dominated modes and before cross-correlating it to the NIRB. The resulting signal, related to the 21-21-NIRB bispectrum, traces the reionisation history and can be detected at high significance by the SKA and SPHEREx \citep[SNR$=6.8$ at $z=9$ when observing for 1000~hrs with SKA-\textit{Low}, see][]{SunLidz_2024}.

\subsubsection{The X-ray background}

As discussed above, we expect the IGM to be heated to temperatures above the CMB, driving the signal from absorption to emission, before the bulk of the EoR.  The sources responsible could include X-ray emission from X-ray binaries, AGNs, and shock heated interstellar medium. %\footnote{Note that the high-redshift component of the CXB comes mainly from hard X-rays, which can easily escape from their host galaxy and travel long distances, while the soft X-ray photons produced by the same sources interact with the neutral atoms of the IGM, imprinting the 21\,cm signal. We do not consider this effect here, but see \citet{Fialkov2017MNRAS.464.3498F,Pochinda2024MNRAS.531.1113P,GesseyJones2024MNRAS.529..519G,Katz2025JCAP...10..047K,Dhandha2025MNRAS.542.2292D,Mirocha2025ApJ...983...54M} for high-redshift constraints from CXB.}. 
However, these sources are hardly detectable in the cosmic X-ray background (CXB) radiation permeating our Universe, as they represent less than a few percents of its mean intensity and spatial fluctuations \citep{MaCiardi_2018}. The cross-correlation of the CXB with the 21\,cm signal helps in isolating the EoR/CD contribution to the CXB and contains information on the properties of the X-ray sources and the timing of these epochs \citep{ShanQin_2009,LiangMao_2016}. This cross signal could be detectable if the X-ray survey is deep enough to remove bright sources with an observed flux $\geq 10^{-17}$~erg/cm$^2$/s, whilst covering a wide field of view \citep{MaCiardi_2018}. In this respect, we expect a cumulative SNR $\gtrsim 1$ when cross-correlating SKA data with observations carried out with the Wide Field Imager (WFI) of the upcoming space telescope \textit{New}Athena \citep{CruiseGuainazzi_2025}.

%%%%%%%%%%%%%%%%%%%%%%%%%%%%%%%%%  CROSS w GALAXIES %%%%%%%%%%%%%%%%%%%%%%%%%%%%%%%%%%%%%%
\subsection{Resolved Galaxies}
\label{sec:3-2}

%Next telco: Monthly since February 25, first Tuesday of month, 1pm CEST

%General points to keep in mind: \\
%Overlap with 3.3 LIM -> coordinate (Caroline), e.g. also on figure 2 \\
%Existing results OR also AA4 / AA*? let's first focus on our forecasts we have for draft

% Brief summary expertise:
% Kana: focus Ha and OIII, IM cross galaxy survey, detectability, Ha and OIII theory calculation for params such as SFR
%Yuxiang: existing forecasts for cross 21cm and emission line galaxies, FRESCO NII (z~6) OIII (z~8) OII (z~10) 
% Shintaro: HSC/PSF detectability predictions, foreground contamination -> forecast/line added?
% Anne: Astreaus and LAE-cross
% Caroline: cross with LAE, Ha, connect to LIM chapter
% Sam: cross with LAE, OIII+Ha, forward modeling signal+noise

This section highlights the prospects for statistical cross-correlation of the 21cm signal and galaxy maps. Galaxy surveys are ideal candidates for cross-correlation with the large-scale 21cm signal due to the potentially wide fields they probe. The range of galaxy survey experiments most promising for detection includes narrow-band dropout, slitless spectroscopy and slit spectroscopic followup. The 21cm-galaxy cross-correlation signal is sensitive to the progress of reionization, its topology, and the heating state of the IGM -- while also encoding ensemble properties of galaxy populations and their environments. Studies have focused on 1) the detectability of the signal depending on survey characteristics and 2) the signal sensitivity to and dependence on reionization parameters and source properties.

% experiment characteristics and detectability
\subsubsection{Survey characteristics and detectability of the 21cm-galaxy cross-correlation}
\begin{figure}[h]
    \centering
    \includegraphics[width=0.2\columnwidth]{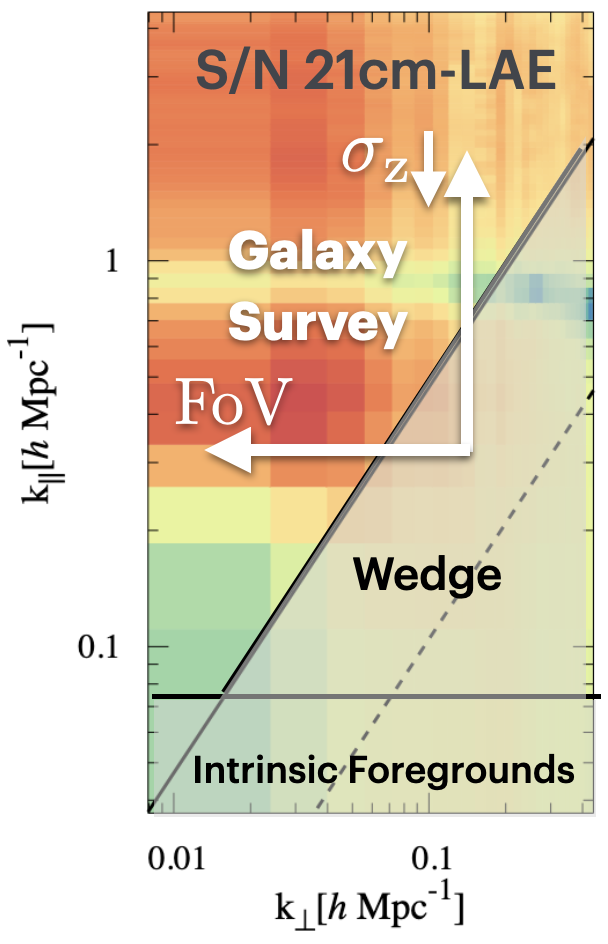}
	   \includegraphics[width=0.78\columnwidth]{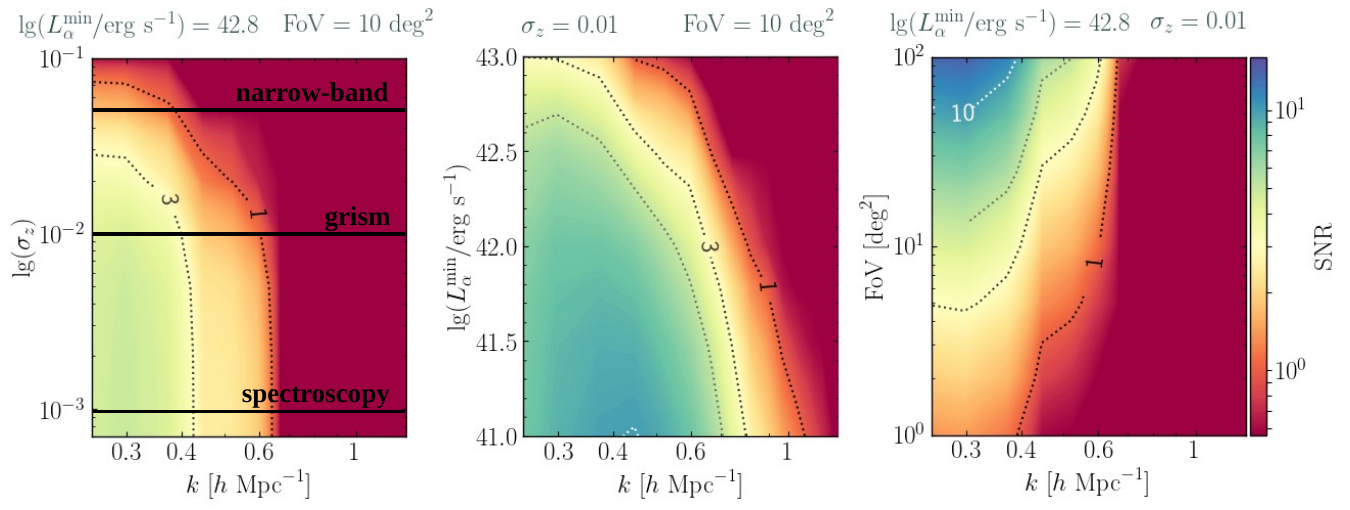}
    \caption{Left panel: Example of the 2D 21cm-galaxy cross-correlation S/N. The signal overlap region in k-space depends on galaxy survey redshift uncertainty and survey area, as well as SKA1-Low foreground assumptions~\citep{Yoshiura_2018};
    Right panels: the S/N as a heatmap for the spherical 1D 21cm-galaxy cross-correlation PS for given survey specifications (y-axis) as a function of k (x-axis), assuming $1000\,$h SKA1-low noise in AA4 configuration~\citep{Heneka_2020,hutter_2023} and optimistic foregrounds. %for now that's what Hutter, Heneka+25 in prep. is based on
    Shown are from left to right the impact on the S/N for a) redshift uncertainty, b) luminosity threshold, c) FoV. 
    }
    \label{fig:3-2-1}
\end{figure}

Emission lines are needed for redshift localization of galaxies as well as to efficiently detect them without expensive spectroscopic campaigns.  Galaxies with strong emission lines are most the most promising candidates, such as  Lyman-alpha emitters (LAE), as well as H-alpha and [OIII] emitters.  % OIII optical/FIR (ALMA)
For cross-correlation with the SKA, several studies have focused on signal shape and detectability of the 21cm-LAE cross-correlation, particularly for LAEs detected via narrow-band dropout with Subaru Hyper-Suprime Cam~\citep[HSC;][]{Sobacchi_2016,Hutter_2017, Kubota_2018, Kubota_2020,Heneka_2020,Gagnon-Hartman_2025}, driven by ongoing measurements of comparably wide fields at redshifts $z\sim 5.7, \, 6.6, \, 7.3$. %relevant for the EoR that are accessible with ground-based observations. 
Fig.~\ref{fig:3-2-1} (left panel) shows a sketch of the dependency of the signal overlap region for the 21cm-galaxy cross-correlation on the galaxy survey redshift uncertainty, $\sigma_{\rm z}$, its field-of-view (FoV), and the 21cm foreground contamination. The smaller the $\sigma_\mathrm{z}$, the larger the FoV, and the less extended the foreground wedge, the larger the signal overlap in cross-correlation.
%On the theory side, as Lyman-alpha is a resonant line damped by neutral hydrogen, LAEs are surrounded by ionized regions and should therefore display a characteristic correlation signal that depends on the progress of reionization.
These studies consistently report that during the EoR, the amplitude of the cross-correlation function, or coefficient, generally decreases (increases) with the ionized (neutral) fraction of the IGM $x_\mathrm{HII}$ ($x_\mathrm{HI}$) and is negatively correlated in the post-heating regime, when the IGM is, on average, hotter than the CMB and spin temperature fluctuations saturate. The strongest (anti-)correlation signal occurs at small scales (< few cMpc) as ionized regions around LAEs appear as cold spots. At large scales the cross-correlation coefficient becomes approximately zero, see Fig.~\ref{fig:3-2-3} (right panel). For fiducial SKA1-Low 1000h observations, both for AA* and AA4 antenna configurations, the anti-correlation at small to intermediate (few tens of cMpc) scales has been reported to be detectable, depending on the assumed galaxy survey characteristics. This makes it possible to track the progress of reionization by distinguishing between different average HI fractions of the IGM. We note that the S/N achievable also critically depends on the foreground model adopted for SKA1-Low.%this seems to be one of the main reaons for lower S/N narrow-band in Gagnon-Hartman+25 as compared to earlier studies, e.g. Heneka_2020 that assumed 21cmSense 'opt' wedge settings
% Even cross-correlation detections with SKA-precursor instruments, such as LOFAR, MWA and HERA, might be within reach jointly with LAE detections by Subaru/HSC narrow-band dropout and Roman grism, again depending on assumed survey characteristics ~\citep{Wiersma_2013,Kubota_2018,LaPlante_2023}.

In Fig.~\ref{fig:3-2-1}, the three right panels show, from left to right, the S/N dependence as a heatmap for the spherical 1D 21cm-galaxy cross-correlation PS, for a range of realistic galaxy survey specifications and noise for $1000\,$h SKA1-low in AA4 configuration. First, we note that across survey characteristics accessible k-modes range from $\sim 0.2 $ to $\sim 1\,$Mpc$^{-1}$. Regarding the dependence on $\sigma_\mathrm{z}$, both slitless (e.g. grism) and slit spectroscopic surveys similarly enable measurements with S/N$\,>3$, and narrow-band surveys reach S/N$\,>1$ across several k-bins for a fiducial $10\,$deg$^{2}$ field, well within reach for for narrow-band and grism observations. As for $\sigma_\mathrm{z}$ we note a relatively mild dependence on the range of luminosity thresholds or survey depths (middle), but a stronger dependence on the survey FoV (right). For slitless and slit spectroscopic FoV$\,\gtrsim$ multiple 100\,arcmin$^{2}$ to $1\,$deg are desirable, whereas narrow-band surveys are expected to require FoV$\,>10\,$deg$^{2}$ for significant detections.

\begin{figure}[h]
    \centering
    \includegraphics[width=0.95\columnwidth]{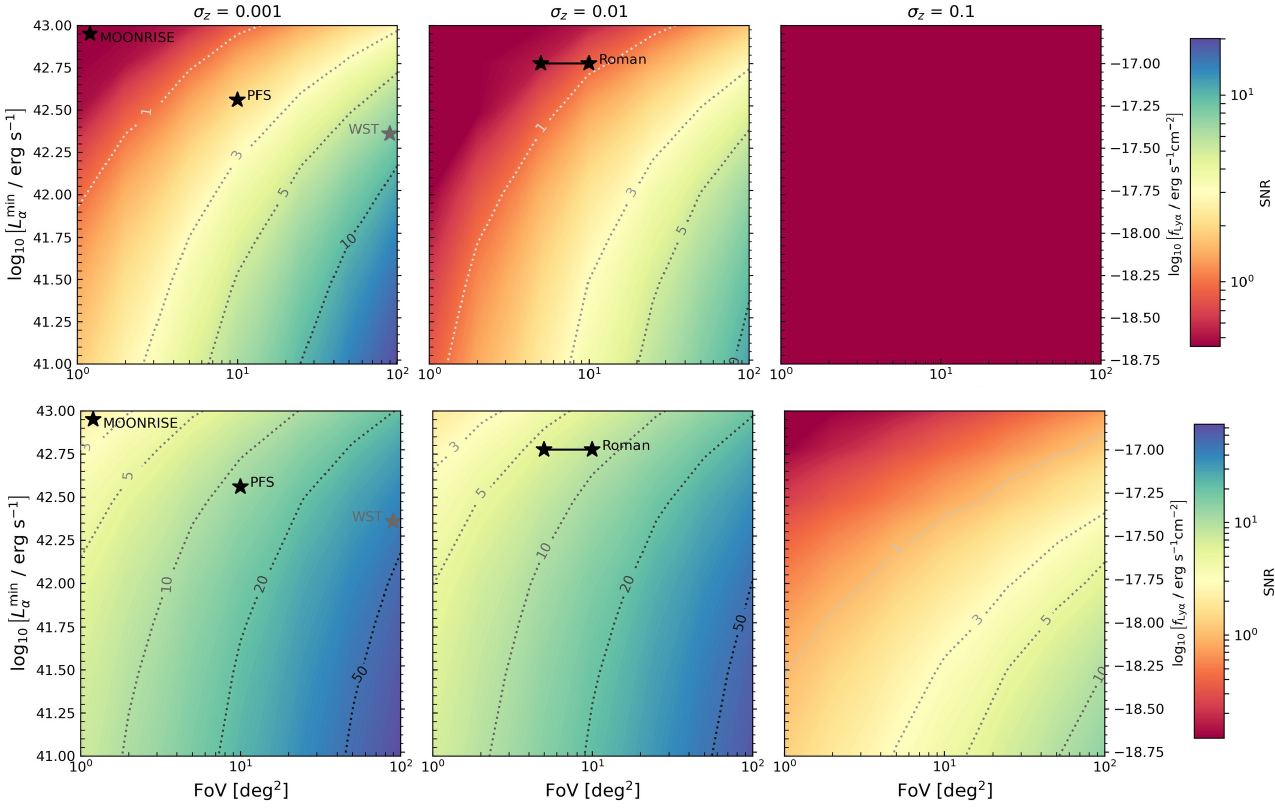}
    \caption{Cumulative SNR of $z\simeq7$ galaxy--21cm cross-power spectrum detection depending on FoV and depth, for slit spectroscopic (left column), grism (middle column), narrow-band (right column) galaxy surveys and two 21cm foreground scenarios, moderate (upper panels) and optimistic (lower panels),
    %\textcolor{blue}{new plot as discussed on July 1st, this version for the moment is AA*, to be finalised, for now plot following Hutter, Heneka in prep. 
    see also~\citet{Gagnon-Hartman_2025}. 
    }
    \label{fig:3-2-2}
\end{figure}

Several studies have extended their analysis to the cross-correlation with the 21cm signal for spectroscopically confirmed LAEs via Subaru Prime Focus Spectrograph (PFS) follow-up, VLT MOONS, and ELT MOSAIC slit spectroscopy, as well as grism observations with the Roman Space Telescope~\citep{Hutter_2018,Vrbanec_2020,Gagnon-Hartman_2025}. Here, grism and slit spectroscopy can significantly enhance detectability and S/N of the cross-signal due to the reduced redshift uncertainty, even at moderate survey depths and survey areas of a few degrees. Besides LAEs, H-alpha ($z<7$), [NII] ($z\sim 6$), H-beta and [OIII] ($z>7$), and [OII] ($z\sim 10$) line-emitting galaxies are also potentially high S/N tracers for the creation of galaxy maps and cross-correlation with the 21cm signal. For example, surveys with JWST NIRCam, using either grism modules or narrow-band filters, are suitable candidates~\citep{Moriwaki_2019}. We note however, that the S/N of the cross-correlation is sensitive to the assumed 21cm foreground model. In Fig.~\ref{fig:3-2-2} we show the S/N dependency for specific galaxy surveys (vertical lines) in two foreground wedge scenarios, a) modes below the horizon plus a $0.1\,h$~Mpc$^{-1}$ buffer are excluded (upper panels) b) all modes larger than the beam are recovered (lower panels).

\subsubsection{The 21cm-galaxy cross-signal as a probe of reionization and source properties}
\label{sec:21cm-galaxy}

For the 21cm-galaxy cross-correlation signal during the EoR, studies often rely on either full hydrodynamical and radiative transfer simulations or semi-numerical simulations coupled with (semi-)analytic and radiative transfer recipes to jointly model both the large-scale 21cm signal and line-emitting galaxies.

\begin{figure}[h]
    \centering
    \includegraphics[width=0.95\columnwidth]{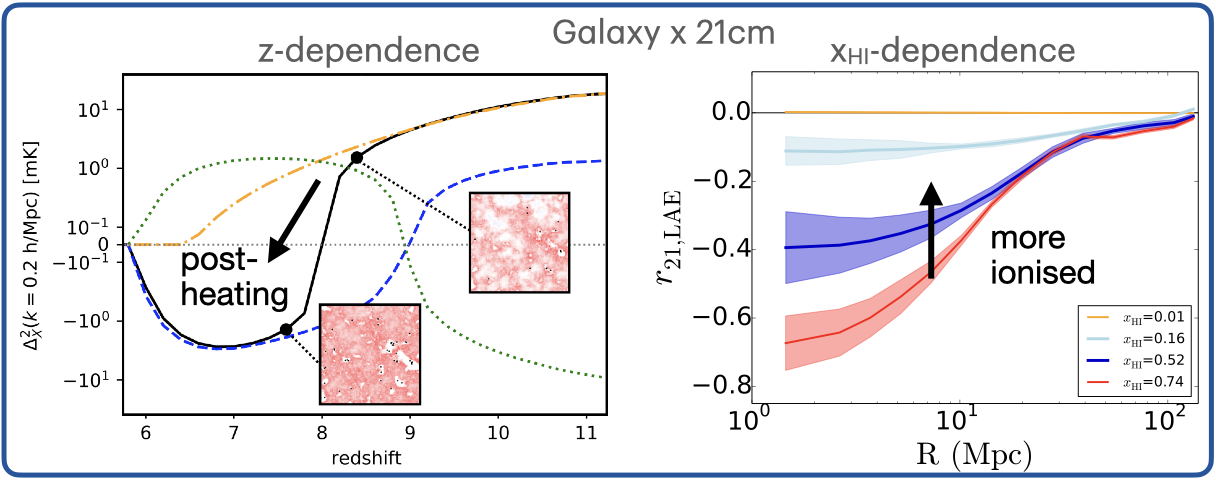}
    \caption{Left panel: The redshift evolution of the cross-power spectra between galaxies and fluctuations in the 21cm intensity (black solid), neutral fraction (blue dashed), gas temperature (orange dash-dotted), and matter density (green dotted) \citep{Moriwaki_2024};
    %Mock example of an observed Lyman-alpha galaxy map at a neutral fraction of $x_\mathrm{HI}\sim 0.5$ (red stars) and the neutral fraction field (dark blue)~\citep{hutter_2023}; 
    % switch to 'theory plot' by Kana https://arxiv.org/abs/2404.0826, Fig. 3 top middle
    right panel: cross-correlation coefficient between galaxies and 21cm signal as a function of distance $R$ for different $x_\mathrm{HI}$ values~\citep{Heneka_2020}. 
    %right panel: Mock example of [OIII] line emitters (points) and the corresponding 21cm brightness map (orange) at $x_\mathrm{HI}\sim 0.5$~\citep{Moriwaki_2019}.
    }
    \label{fig:3-2-3}
\end{figure}

Across modeling approaches, studies agree that the 21cm-galaxy cross-correlation function is sensitive to HI fraction and distribution in the IGM, with its amplitude generally depending on the progress of reionization and its shape serving as a measure of the reionization morphology (i.e., the size and distribution of ionized regions); examples for modeled LAE and [OIII] emitter x 21cm fields, as well as the signal dependence on the HI fraction are shown in Fig.~\ref{fig:3-2-3}. Additional information is encoded on the reionization morphology; the cross-correlation amplitude at small scales decreases with increasing correlation between the density field and the ionizing emissivity of the galaxy population as traced by ionization. This means that the small-scale amplitude is reduced if lower-mass galaxies drive reionization. The sign and amplitude of the cross-correlation coefficient also depend on the temperature of the neutral IGM (Fig.~\ref{fig:3-2-2}, left panel). Here, an IGM colder than the CMB results in a positive correlation, while in the post-heating regime (saturated spin temperature), the correlation becomes negative, as shown for LAEs in~\citet{Heneka_2020} and [OIII] emitters in~\citet{Moriwaki_2019}. Finally, the cross-correlation, like the galaxy auto-correlation, depends on ensemble properties of galaxies -- such as luminosities (where the resulting correlation signal is determined by the luminosity threshold or depth of the survey) and radiative escape through surrounding media, linking intrinsic luminosities associated with star formation rates to observable luminosities.

For LAEs in particular, predictions of the 21cm-LAE cross-correlation from different modeling approaches can be mapped to a common analytical form, exploiting the fact that LAEs are located in sufficiently large ionized regions that allow the Lyman-alpha line to redshift out of resonant absorption. In detail, the small-scale (< few cMpc) cross-correlation amplitude directly scales with both the average HI fraction in the IGM and the spin-temperature-weighted overdensity in neutral regions. While previous studies have also noted that the typical size of ionized regions is related to the inversion point of the cross-correlation function, in~\citet{Hutter_2023b}, the distribution of ionized region sizes and its cumulative distribution function, corresponding to the average profile of the HI fraction around LAEs, was shown to determine the shape and inversion point of the 21cm-LAE cross-correlation.

%%%%%%%%%%%%%%%%%%%%%%%%%%%%%%%%%  CROSS w LIM %%%%%%%%%%%%%%%%%%%%%%%%%%%%%%%%%%%%%%%
\subsection{Multi-wavelength Line Intensity Mapping (LIM)}
The bright spectral line emissions from the galaxies, such as [CII] 158$\mu$m~\citep{Gong+2012, Silva+2015, Yue+2015, Lagache+2018, Yue+2019, Sun+2021, Bethermin+2022, Dizgah+2022, Yang+2022, Kannan+2022, Karoumpis+2022, Padmanabhan+2022, Murmu+2021, Murmu+2022, Murmu+2023, Sun+2023, Van-Cuyck+2023, Karoumpis+2024, Liang+2024, Liu+2024, Casavecchia+2025}, CO~\citep{Carilli+2011, Lidz+2011, Mashian+2015, Breysse+2017, Breysse+2022, Chung+2023, Dosibhatla+2025}, Ly-$\alpha$ and H-$\alpha$~\citep{Silva+2013, Pullen+2014, Heneka+2017, Silva+2018, Visbal_&_McQuinn+2018, Chang+2019, Heneka+2021, Kannan+2022, Mas-Ribas+2023, Sun+2023, Padmanabhan+2024, Lee+2025}, [OIII] 88$\mu$m~\citep{Padmanabhan+2022}, etc., can be good candidates to conduct galaxy LIM, which detects the integrated flux from the numerous unresolved galaxies within  voxels. It is therefore similar to the concept of intensity mapping of 21cm emission, originating from the diffuse IGM during the EoR~\citep{Furlanetto+2004, Bharadwaj+2005}, with the difference that the sources of line-emission are discrete in nature (galaxies). LIM has the advantage that it can probe  large volumes of the Universe with accurate redshift localization, compared to the traditional galaxy surveys. This implies that galaxy LIM surveys are ideally suited for synergies with large-scale 21cm intensity mapping surveys  for which foregrounds contaminate transverse modes making overlapping redshift footprints essential~\citep{Lidz+2011, Gong+2012, Heneka+2017, Murmu+2021, Murmu+2022, Murmu+2023, Kannan+2022, Sun+2021, Sun+2023, Fronenberg+2024}. 

%\\ \textcolor{blue}{[CH: if Lya is included in the above, I would phrase this slightly more general - as Lya also can probe the CGM to IGM scales and thus is not only 'galaxy' LIM or LIM of discrete sources necessarily (but it is mostly, I agree).]}

\subsubsection{Motivation for multi-tracer LIM}
\label{sec:3-3-1}
 Intensity mapping of the EoR using the redshifted 21cm signal is essentially a probe of the IGM and its evolution with cosmic time. However, 21cm intensity maps do not provide a direct probe of the sources of reionisation. Thus, a direct probe of the sources of reionisation, the majority of which are expected to be extremely faint galaxies and thus cannot be detected via the usual galaxy surveys, is essential to obtain a complete picture of the EoR. Without a direct probe of the sources of EoR, one will face challenges in constraining the EoR with 21cm maps alone, due to the inherent degeneracies in reionisation parameters ~(e.g. \citealt{Greig2015MNRAS.449.4246G, Ghara+2021}). Here, LIM of galaxies through their bright spectral lines comes to our rescue as they complement the IGM 21cm maps and thus help in breaking the degeneracies in reionisation model parameters.
 
 Going from smaller to tentatively larger scales probed by multi-wavelength LIM, the H-$\alpha$ line probes the SFR integrated over galactic luminosities of stellar origin, but can be obscured by dust. In optical wavelength, the [OIII] line probes emission from star formation as a tracer of HII regions~\citep{LIM:2017}. The [CII] $158\mu$m line observed in FIR originates from various phases of the ISM within the galaxies, such as the molecular phase, atomic, and ionised gas~\citep{De_Looze+2014, Leung+2020}. Thus [CII] is also a tracer of the total star formation rate. In the cm/mm to sub-mm regime, the [CO] line emissions arise from the rotational transitions of the CO molecule and they serve as a probe of the molecular gas content in galaxies, which fuels star formation~\citep{Carilli_&_Walter+2013}. Finally, the Ly-$\alpha$ rest-frame UV line potentially probes a multitude of scales; its galactic contribution probes the emission from star formation, and it extends into the CGM through scattering and recombination as a tracer of ionised and partially ionised regions, and even to the IGM -- even though at largely decreased intensity -- for example due to down-scattering of Ly-n emission. These line emissions, therefore, are probes of different aspects of star-formation, radiative escape, as well as state and properties of gaseous media from the ISM, over CGM, to the IGM, and thus help us shed light on cosmic reionisation.
Therefore, multi-wavelength LIM will be essential to probe the EoR.

\subsubsection{Multi-wavelength LIM experiments}
\label{sec:3-3-2}
\begin{itemize}
\begin{comment}
    \item \textbf{CONCERTO}: The CarbON CII line in the post-rEionization and ReionizaTiOn (CONCERTO) is an instrument installed on the APEX telescope, and designed to conduct LIM with the redshifted [CII] $158\mu$m line, within the frequency range of approximately $\sim$ $130 - 310$ GHz~\citep{CONCERTO_collaboration+2020}. Since the [CII] $158\mu$m line is known to correlate well with the star-formation within the galaxies, this survey aims to map the 3D fluctuations of this line emission on a large scale, while aiming to answer how star-formation is related to galaxy evolution and the role of [CII] $158\mu$m emitters in cosmic reionization. \citet{CONCERTO_collaboration+2020} has made forecasts on the detectability of [CII] $158\mu$m auto power spectrum using models from~\cite{Serra+2016}, with 1200 hrs of observation and redshift width of $\Delta z \sim 0.6$. Their pessimistic scenario suggests an SNR between $1.1$ - $4.3$ for the [CII] auto power spectrum at $z \approx 6.2$.
\end{comment}
    \item \textbf{FYST}: The Fred Young Submillimetre Telescope~\citep[FYST,][]{CCAT-Prime_Collaboration+2023} is well suited to conduct LIM of EoR using the [CII] $158\mu$m line emission from galaxies. The survey plan suggested by~\citet{CCAT-Prime_Collaboration+2023}, relevant for EoR science, can be a survey of two $\sim 4$ deg$^2$ fields, one being the Extended-COSMOS~\citep[E-COSMOS,][]{Aihara+2018} and the other being the Extended Chandra Deep Field South~\citep[E-CDFS,][]{Lehmer+2005}, including the Hubble Ultra Deep Field~\citep[HUDF,][]{Beckwith+2006}, with an observation time of $\sim 2000$ hrs for each field. FYST will have an Epoch of Reionization Spectrometer (EoR-Spec) module, covering frequencies between $210$ and $420$ GHz and optimized to probe [CII] $158\mu$m emission over redshifts $3.53 \lesssim z \lesssim 8.05$~\citep{Cothard+2020}.
    ~\citet{Roy+2023} forecasted a cumulative SNR of $4$ at $z \sim 7.6$ in their pessimistic scenario for the EoR-Spec of FYST. On the other hand,~\cite{Clarke+2024} have predicted lower limits on the [CII] $158 \mu$m power spectrum for the EoR-Spec using the COSMOS 2020 galaxy catalogue data~\citep{Weaver+2022}, which are $\Delta^2 (k = 1\, {\rm Mpc}^{-1}, 5.34 < z < 6.31) = 9.8 \times 10^5\, ({\rm Jy/sr})^2$and $\Delta^2 (k = 1\, {\rm Mpc}^{-1}, 6.75 < z < 8.27) = 2.77 \times 10^5\, ({\rm Jy/sr})^2$.~Estimations from works such as~\cite{Murmu+2021, Karoumpis+2022, Roy+2025b} are consistent with the $\Delta^2_{\rm CII}(k)$ lower limit, placed at $6.75 < z < 8.27$ and $k = 1$ Mpc$^{-1}$. However, models from~\cite{Kannan+2022} and~\cite{Sun+2023} are not consistent with these lower limits, where these models predict a lower value of $\Delta^2_{\rm CII}(k)$, falling in the redshift interval of $5.34 < z < 6.31$ and $6.75 < z < 8.27$, at $k = 1$ Mpc$^{-1}$. To be more precise,~\cite{Kannan+2022} admits that due to the lack of adequate modelling of [CII] $158 \mu$m emission, their predictions of the line luminosity fall below those from works such as~\cite{Lagache+2018} and~\cite{Leung+2020}.
    \item \textbf{TIME}: The Tomographic Intensity Mapping Experiment (TIME) is an imaging spectrometer array that has a wide bandwidth, allowing it to map [CII] $158 \mu$m line emission between redshifts of $6 < z < 9$~\citep{Hunacek+2018, Cheng+2020}. With an initial planned observation time of $1000$ hrs, it started operation in $2021$.~\cite{Sun+2021} has forecasted that detection of the [CII] $158 \mu$m power spectrum is possible with SNR $> 5$.
    \item \textbf{COMAP}: The CO Mapping Array Project (COMAP) is designed to detect the CO line emissions from the early Universe~\citep{Cleary+2022}. It is comprised of a pathfinder instrument having a spectrometer receiver with $19$ feeds installed on a $10.4$ m diameter dish. The pathfinder instrument can operate between $26 - 34$ GHz band, thereby targeting the CO(1-0) line from redshifts of $2.4 \lesssim z \lesssim 3.4$ as well as being able to probe the EoR redshifts within a redshift range of $5.8 \lesssim z \lesssim 7.8$ via the CO(2-1) line emission. Therefore, naturally, the CO(1-0) line acts as an interloper for the CO(2-1) line emission from the EoR. Apart from this, the COMAP-EoR instrument, planned to operate in the $12 - 20$ GHz band, will be sensitive to the CO(1-0) emission out to the EoR. The observation is planned based on three $\sim 4$ deg$^2$ fields having overlap with HETDEX fall and spring fields~\citep{Gebhardt+2021}. Currently, most of the analysis is focused on extracting the CO(1-0) signal from the post-EoR regime, with the latest work constraining the CO(1-0) power spectrum to $kP_{\rm CO}(k) < 2400 - 4900\,\mu$K$^2$ Mpc$^2$ in the $k$ range of $0.09 < k < 0.73$ Mpc$^{-1}$, and in the redshift range of $2.4 - 3.4$, with $95$ percent confidence interval~\citep{Stutzer+2024}. Although currently, not many works are available showing numerical estimates of the CO(1-0) power spectrum at $2.4 \lesssim z \lesssim 3.4$, the optimistic models from~\cite{Padmanabhan+2018} suggest that, within their uncertainty range, the maximum predicted $kP_{\rm CO}(k)$ is approximately $\sim 1.6\times10^3$ $\mu$K$^2$ Mpc$^2$ at $k\sim0.7$ Mpc$^{-1}$, for $z=3$. On the other hand,~\cite{Sato-Polito+2023} suggest that their model predicts CO(1-0) power spectrum at $kP_{\rm CO}(k) \sim 360$ $\mu$K$^2$ Mpc$^2$ at $k\sim0.09$ Mpc$^{-1}$ and $kP_{\rm CO}(k) \sim 10^3$ $\mu$K$^2$ Mpc$^2$ at $k\sim0.7$ Mpc$^{-1}$ for $z=2.8$. This suggests that the COMAP upper limits at these redshifts are not quite stringent yet. However, both the CO(2-1) and the CO(1-0) signal from the EoR regime still remains to be detected.
    \item \textbf{SPHEREx}: The Spectro-Photometer for the History of the Universe, Epoch of Reionization, and Ices Explorer~\citep[SPHEREx,][]{SPHEREx:2014,SPHEREx:2020} is a NASA satellite mission launched in 2025 designed specifically for multi-line intensity mapping to perform an all-sky spectro-photometric survey in the near-infrared. It provides spectroscopic imaging across $0.75$ to $5$ micron bands in $102$ frequency channels at $6.2$$''$ spatial resolution, enabling detection of key emission lines such as H-$\alpha$, H-$\beta$, [OII], [OIII], and Ly-$\alpha$. Observing these lines across a broad redshift range allows for tracing star formation and galaxy evolution over cosmic time. During the EoR for redshift-overlap and potential cross-correlation with the 21cm background, the mission maps the line intensity of H-$\alpha$, Ly-$\alpha$, and potentially [NII], up to $z\sim 6-8$ depending on the line. 
\end{itemize}

\subsubsection{Challenges for LIM}
\label{sec:3-3-3}

Although the LIM technique can allow for a survey of large cosmological volumes and probe the EoR, it faces certain challenges, one of them being the issue of foreground/interlopers. Given an observation over a certain wavelength range, this band can contain both the signal originating from the target redshift of our interest, and also contaminations from sources at different redshifts. In the case of the 21 cm signal from the EoR, it is known that foregrounds will severely contaminate it \citep{DiMatteo+2002,Santos+2005,Ali+2008}, typically by orders of magnitude in terms of signal strength. The sources of foreground in this case could be from galactic synchrotron emission as well as extragalactic radio point sources. On the other hand, for LIM signals like [CII] $158\mu$m signal originating from the EoR, the CO lines originating from galaxies at much lower redshifts act as interlopers \citep{Gong+2012,Silva+2015}. 

\subsubsection{Synergies of 21cm and other multi-wavelength LIMs for probing the EoR }
%\\ \textcolor{blue}{Or maybe: Synergies of the 21cm and further LIM lines for probing the EoR}}
\label{sec:3-3-4}
Cross-correlation of the 21cm observations from IGM with intensity maps of various emission lines from galaxies (originating from the same patch of the sky) e.g. [CO], [CII], [OIII], H-$\alpha$, and Ly-$\alpha$ introduced in Sec.~\ref{sec:3-3-1} provides a powerful probe of the reionization process. These cross-correlations are sensitive to the intrinsic properties of the emitters, such as their luminosity-halo mass relation and radiative escape fractions, as well as to CGM/IGM properties. Further, cross-correlation between two independent LIM fields (e.g. 21cm from IGM and [CII] or [CO] from galaxies) observed at different frequencies but originating from the same redshift opens up a unique window to mitigate the effect of foregrounds. The signals from these two LIM fields, originating from the same redshift and from the same piece of sky, are expected to be correlated (positively or negatively, depending on the stage of reionisation and length scale of observation), but their foreground and noise are expected to be uncorrelated and thus will not contribute to the signal cross-power \citep{Lidz+2011, Carilli+2011}. However, foreground removal or avoidance is still necessary, as the residual foregrounds will increase the uncertainty in the estimation of the cross-power \citep{Fronenberg+2024}. 

The cross-power spectrum of [CII]-21cm lines from observations by next-generation experiments e.g. FYST and SKA will be detectable with SNR $\geq 10$ \citep{Dumitru+2019, Karoumpis+2022, Karoumpis+2024}. Such detection, first of all, will act as a confirmatory test for the detection of the cosmic signals from the EoR. Further, the cross-power spectrum will be capable of putting tighter constraints on the minimum halo mass capable of producing Lyman-C photons that can escape galaxies, better
than 21cm measurements alone. Additionally, one would be able to constrain the end of reionisation redshift from the sign flip of cross-power at large scales more precisely. However, while inferring the cosmology and astrophysics from the [CII]-21cm cross-power, one should take into account the impact of light-cone effect on this signal statistics, as it can change the cross-power amplitude by $\approx 20\%$ or more at large length scales ($k \leq 0.1\, {\rm Mpc}^{-1}$) \citep{Murmu+2021}. As both of these LIM signals, originating from IGM and galaxies, are inherently non-Gaussian in nature, one can quantify their non-Gaussianity by estimating the cross-bispectrum between them. \citet{Beane+2018} demonstrated that it would be possible to estimate the redshift evolution of the product of the mean 21cm brightness temperature and 21cm bias using the 21cm-[CII]-[CII] cross-bispectrum. The same can be done with 21cm-Lyman-$\alpha$ cross-bispectrum using the observations from SPHEREx and HERA, as both have overlapping sky coverage \citep{SPHEREx:2014, SPHEREx:2020,HERACollaboration2023,HERA+2017}. The cross-correlation between the 21cm observation of IGM with the SKA and any other LIM observation from galaxies via FYST, TIME, COMAP and SPHEREx-like instruments has the potential to constrain reionisation history, source models, ionisation morphology and IGM physics in a more robust fashion compared to the auto-correlation of any of these signals alone.

%\begin{figure}[h]
%    \centering
%    \includegraphics[width=\columnwidth]{Plots/Chapter_3-3/LIM_21cm_Lya_example.png}
%    \caption{Left to right: figure example, from LIM simulation or mock, over cross-power spectrum or correlation (arXiv:1611.09682), to S/N estimates (arXiv:2104.12739).
%    }
%    \label{fig:3-3-1}
%\end{figure}
%\subsubsection{Outlook:}

%%%%%%%%%%%%%%%%%%%%%%%%%%%%%%%%%%%%%%%%%%%%%%%%%%%%%%%%%%%%%%%%%%%%%%%%%%%%%%%%%
%%%%%%%%%%%%%%%%%%%%%%%%%%%%%%%%%  IMAGES %%%%%%%%%%%%%%%%%%%%%%%%%%%%%%%%%%%%%%%
\section{Image-based synergies}

In the previous section we focused on statistical cross-correlations, as will be relevant for initial SKA observations.  Eventually with Phase 2, we will be able to study images of individual regions, nailing down the galaxy/AGN -- IGM connection.  Here we discuss possibilities for such image based synergies.

%%%%%%%%%%%%%%%%%%%%%%%%%%%%%%%%%  IMAGES / galaxies %%%%%%%%%%%%%%%%%%%%%%%%%%%%%%%%%%%%%
\subsection{Galaxies}
\label{sec:4-1}

This subsection describes methods that combine the 21cm signal and galaxy data by directly imaging the 21cm signal around galaxies, including quasars, rather than using cross-correlations. These methods trace quantities similar to those from 21cm-galaxy cross-correlations, such as the sizes of ionised regions and the HI fraction of the IGM, but allow for a more detailed examination of specific environments and galaxy populations. Broadly, they can be divided into two categories: (1) mapping ionised regions around individual galaxies and (2) stacking galaxy-informed 21cm images.

\subsubsection{Constraints from mapping ionised regions around galaxies}

\begin{figure}[h]
    \centering
    \includegraphics[width=0.9\columnwidth]{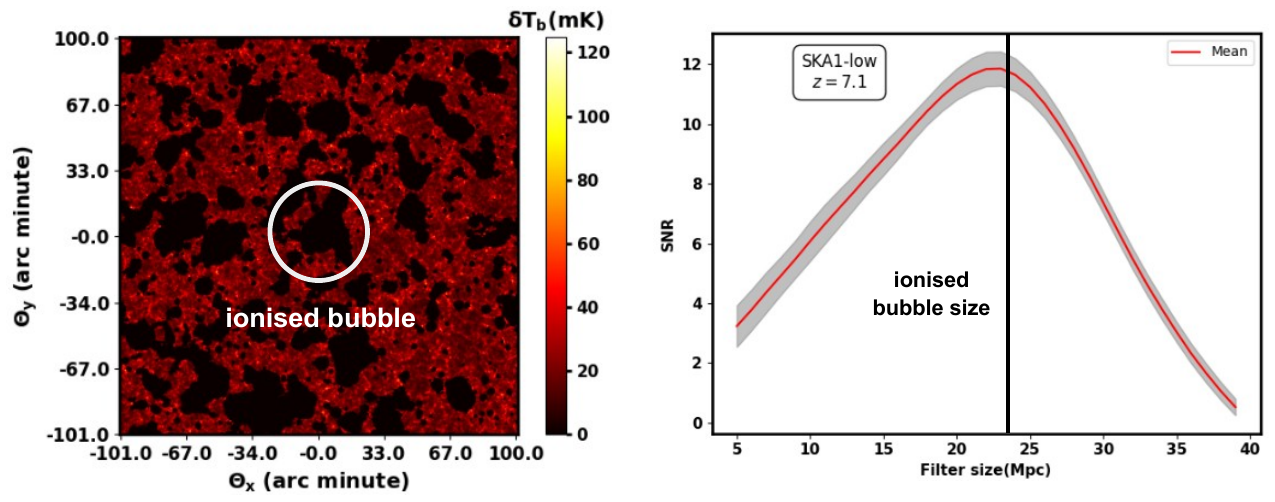}
    \caption{{\it Left:} 21cm differential brightness temperature distribution at $z=7.1$ and $\langle \chi_\mathrm{HI} \rangle=0.52$ centering on an ionised regions with a size of $23.5$~cMpc \citep{mishra25}. {\it Right:} Signal-to-noise ratio from matched filtering applied to 20 independent noise realizations of 200 hours SKA1-Low observations \citep{mishra25}. The shaded region indicates the $1\sigma$ uncertainty from the realizations.}
    \label{fig:matched_filtering_method}
\end{figure}

The matched filter method, a signal processing technique designed to extract signals from noisy data, could be a promising technique for studying 21cm images \citep{datta07,datta08,majumdar11,majumdar12,mishra25}. This method enhances the detection of ionised regions by convolving a filter, shaped to match the expected 21cm signal, with the observed visibilities. The signal-to-noise ratio (SNR) of this convolution is maximised when the filter closely matches the true 21cm signal (c.f. Fig.~\ref{fig:matched_filtering_method}). By testing different spherical ionised region sizes, this approach effectively determines ionised region sizes and traces the IGM HI fraction.

Studies have explored filters mimicking a spherical ionised region within a uniform HI background, using toy models and semi-numerical simulations to assess the feasibility of measuring the ionised region sizes (or bubbles) around galaxies, in particular those hosting quasars, with SKA1-Low. For example, \citet{mishra25} found that with 100–200 hours of SKA1-Low observations, ionised regions larger than $\sim20$~cMpc around galaxies could be detected in a Universe that is up to 50\% ionised. 
%Early JWST observations of Lyman-$\alpha$ emitters have already uncovered an ionized bubble of the required size at $z\approx 8$ \citep{Witstok25}, and 
 The combination of wide-area surveys for high-redshift Lyman-$\alpha$ emitters in regions of the sky suitable for SKA observations should, when followed-up with deep JWST spectroscopy (e.g. \citealt{ Lu24, Nikolic25}), result in a sample of ionized bubbles suitable for size measurements based on 21cm observations. A key advantage of targeting ionised regions around already detected galaxies is that determining their sizes is primarily limited by the instrument's resolution. In contrast, blind searches may struggle to detect smaller ($<10$~cMpc) ionised regions.
 %due to increasing fluctuations in the cosmic HI 21cm signal as the IGM HI fraction decreases \citep{datta08}. 
  Since the 21cm signal amplitude around the ionised region is proportional to the mass-averaged HI fraction outside the ionised region, the SNR value can also be used to trace the IGM’s mean HI fraction \citep{mishra25}. While promising, further work is needed to fully understand how ionised region detection and HI fraction constraints will behave during the early cosmic epochs, when the first ionised regions begin to form, the IGM HI fraction is high, and IGM heating is still ongoing. A different approach delineates ionised bubble boundaries in three dimensions, enabling robust inference of quasar lifetimes, ionizing luminosities, and IGM conditions. \citet{Bolgar2018} illustrated how variations in quasar duty cycles leave distinct signatures in 21\,cm tomography, though detectability remains challenging.

\subsubsection{Constraints from galaxy-informed stacking of 21cm fields}

A few studies have explored how stacking 21cm fields around pre-selected galaxies or empty regions can probe ionised region sizes, the global IGM ionisation state, and the reionisation morphology.

\begin{figure}[h]
    \centering
    \begin{minipage}{0.45\textwidth}
        \includegraphics[width=1.\columnwidth]{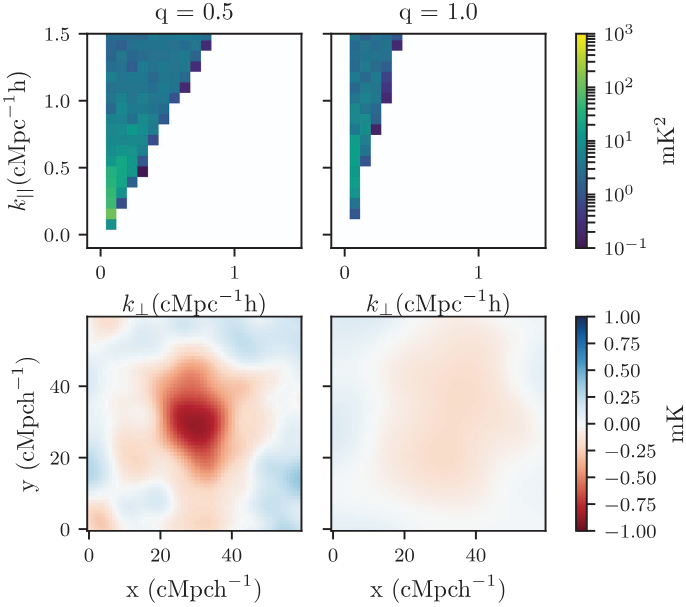}
    \end{minipage}
    \hfill
    \begin{minipage}{0.45\textwidth}
        \includegraphics[width=1.\columnwidth]{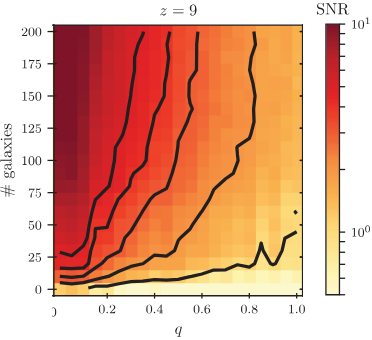}
    \end{minipage}
    \caption{{\it Left:} Stacked simulated 21cm images of ionised regions around galaxies at $z=9$ for two foreground models from \citet{Davies_2021}: maximal contamination ($q=1$) and a reduced case where foreground subtraction halves the horizon slope ($q=0.5$). Top panels show the stacked 21cm signal in cylindrically average Fourier space, and the bottom panels show slices of the centre of the stacked profiles. {\it Right:} Signal-to-noise ratio (SNR) for detecting the stacked ionized regions at $z=9$ as a function of foreground level and number of galaxies. Contours from right to left correspond to SNR = (1, 2, 3, 4, 5).}
    \label{fig:stacking_21cm_images_around_galaxies}
\end{figure}

\paragraph{Stacking 21cm images around galaxies} 
Sources detected by JWST, Roman, or Euclid will reveal average 21cm profiles for different galaxy samples, such as those grouped by ultraviolet (UV) luminosity. \citet{Davies_2021} demonstrated that both the amplitude and width of these profiles increase with the size of ionised regions, with the latter reflecting the greater asymmetry of more spatially extended ionised regions. Additionally, the profile’s sign indicates whether the 21cm signal is in emission or absorption, being negative (positive) when the spin temperature is saturated (unsaturated). 
Analyses of reionisation simulations reveal that in an ideal case without foreground contamination a $5\sigma$ detection at $z\sim9$ is possible with images from 30 galaxies. However, in presence of foregrounds, only images around 100 of the brightest galaxies in the  SKA1-Low FoV could yield a $1.4\sigma$ detection at $z\sim9$; partial foreground removal can achieve detections with fewer images (see Fig. \ref{fig:stacking_21cm_images_around_galaxies}).
By analysing the amplitudes and widths of average 21cm profiles around galaxies of varying UV luminosities, we can constrain the size distribution of ionised regions and the overall ionisation morphology \citep{Wyithe2005,Geil_2017, Davies_2021}. Broader, shallower profiles around UV-bright galaxies --suggesting more asymmetric ionised regions \citep{hutter_2023} -- may indicate reionisation scenarios driven by faint galaxies, though further validation is needed.

\paragraph{Differences in the average 21cm differential brightness temperatures between regions with and without Lyman-alpha emitting galaxies (LAEs)}
\begin{wrapfigure}{r}{0.4\textwidth}
  \centering
  \includegraphics[width=0.38\textwidth]{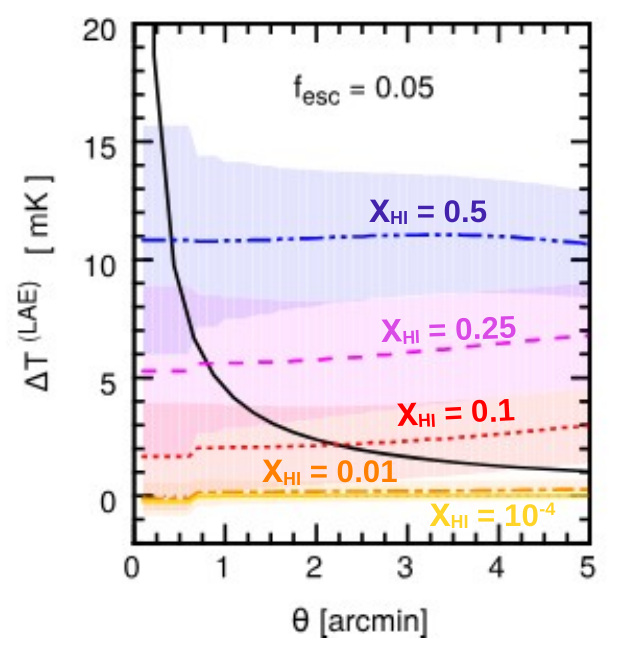}
  \caption{Difference in the 21cm signal between regions with and without LAEs, averaged over $10$ fields, as a function of angular size for 1000 hours SKA observations \citep{Hutter_2017}.}
  \label{fig:diffTb_LAE_selected_21cm_fields}
\end{wrapfigure}
can constrain the IGM’s ionisation state and reionisation morphology. Lyman-alpha radiation is sensitive to HI absorption.  However galaxies are biased tracers of the matter field and so trace regions that are likely overionized/overheated compared to average regions. 
% Therefore, regions with LAE overdensities contain complex and physics-rich information about the ionization state of the IGM.
%have a complex and physics-rich Since Lyman-alpha radiation is highly sensitive to HI absorption, LAEs, which simulations suggest are among the most massive galaxies, serve as tracers of ionised regions. Consequently, the average 21cm differential brightness temperature in regions containing LAEs ($T_\mathrm{LAE}$) is expected to be higher than in regions without them ($T_\mathrm{noLAE}$), with the 21cm signal’s amplitude decreasing in both regions as the Universe becomes more ionised. 
 By measuring the difference between the average 21cm signal ($\Delta T^\mathrm{(LAE)}$) on 1-5~arcmin scales in regions containing and lacking LAEs with 1000 hours SKA observations (V4A array configuration), we can learn about the ionization state and morphology of the IGM.
%robustly distinguish between different IGM ionisation states. Specifically, 
For example, analysing 10 fields with and without LAEs, this method can differentiate an IGM 10\% neutral from one that is 50\% neutral \citep{Hutter_2017}.%, and test our predictions of reionisation proceeding inside-out, where ionised regions percolate from overdense to underdense regions. 
 The measured difference mimics the small-scale 21cm-LAE cross-correlation amplitude and inherits its dependencies on the ionisation morphology. As reionisation is driven more by fainter galaxies, the ionisation field becomes more strongly correlated with the underlying density field, reducing the 21cm signal amplitude difference between LAE and non-LAE regions \citep{hutter_2023}.

% %%%%%%%%%%%%%%%%%%%%%%%%%%%%%%%%%  IMAGES / QSP %%%%%%%%%%%%%%%%%%%%%%%%%%%%%%%%%%%%%
% \subsection{Quasars}
% {\bf lead: Tirth} potential contributors: Pratika, Suman, Kanan, Tirth…

% \textbf{TRC: Several of the topics proposed here are already covered in Section 4.1 (Galaxies). We can either merge this section with 4.1, or alternatively, we need to carefully distinguish and extract the results from 4.1 that are specifically relevant to quasars.}

% Tentative proposed plan:

% \begin{itemize}
% \item Features of ionized regions around quasars, expected 21\,cm signal, optimal redshift, challenges in detecting them.
% \item Theoretical models/simulations, what can we learn about quasars (luminosity, lifetime, duty cycle)  and the IGM (ionization state, temperature)?
% \item Visibility-based detection: matched filtering  
% \item Image-based detection: stacking
% \item Other techniques?
% \item Cross-correlation of ionized bubbled around quasars with Lyman-$\alpha$ spectra 

% \item \textbf{[TRC: Where does 21\,cm forest fit?]}
% \end{itemize}

%%%%%%%%%%%%%%%%%%%%%%%%%%%%%%%%%%%%%%%%%%%%%%%%%%%%%%%%%%%%%%%%%%%%%%%%%%%%%%%%%
%%%%%%%%%%%%%%%%%%%%%%%%%%%%%%%%%  CONC %%%%%%%%%%%%%%%%%%%%%%%%%%%%%%%%%%%%%%%
\section{Conclusions}

In this chapter we reviewed the current state of knowledge about the IGM and galaxies during the EoR.  As we approach the first 21cm detection of the EoR with the SKA, synergies with other facilities will be essential both to verify its cosmological origin as well as to help mitigate unknown systematics.  Therefore it is essential to plan SKA-low observations keeping in mind overlap with other surveys.

\begin{figure}[h]
   \centering
	\includegraphics[width=\columnwidth]{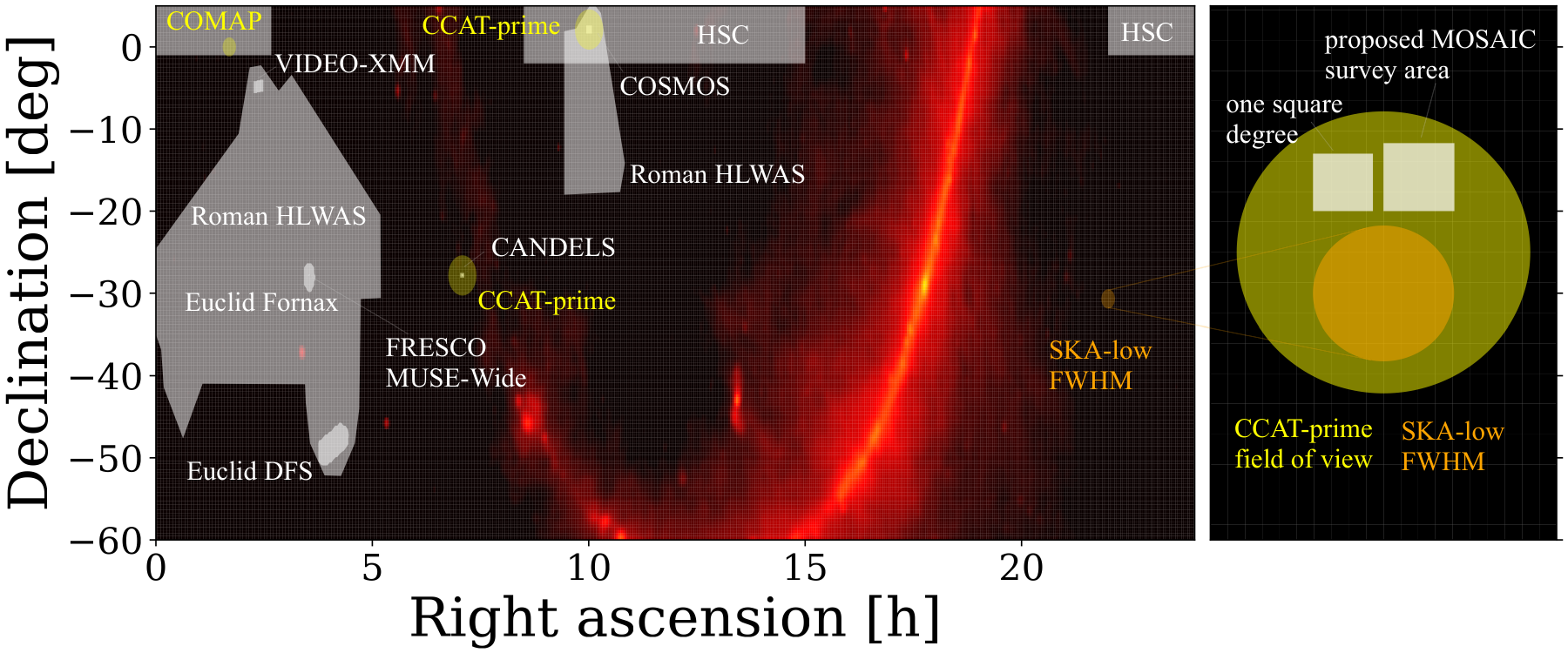}
   \caption{The southern sky from Declination $-60$ to the equator with overlaid galaxy surveys (white) and line intensity mapping surveys (yellow) which have already been carried out or will be carried out by 2030. The SKA-low FWHM is shown in orange for scale, with the right panel comparing the size of a proposed MOSAIC survey to pointings of the SKA-low and CCAT-prime (MOSAIC \cite{2024Msngr.193...24M,Gagnon-Hartman_2025}; CCAT-prime \cite{CCAT-Prime_Collaboration+2023}). The background image was generated using the Global Sky Model at $150$ MHz \citep{2008MNRAS.388..247D}.}
   \label{fig:skymap}
\end{figure}

In Figure \ref{fig:skymap} we show a map of the southern sky with existing and planned high-$z$ galaxy surveys (VIDEO-XMM: \cite{2013MNRAS.428.1281J}; Roman HLWAS-Medium: \cite{2025arXiv250510574Z}; HSC: \cite{2018PASJ...70S...4A}; Euclid Fornax, Euclid DFS: \cite{2025A&A...697A...1E}; FRESCO: \cite{2023MNRAS.525.2864O}; MUSE-Wide: \cite{2017A&A...606A..12H}; CANDELS: \cite{2011ApJS..197...35G}; COSMOS: \cite{2025arXiv250603243S}) overlaid in white, and existing and planned line intensity mapping surveys (COMAP: \cite{Cleary+2022}; CCAT-prime: \cite{CCAT-Prime_Collaboration+2023}) in yellow. Note that the FRESCO and MUSE-Wide surveys are located in the GOODS-S field. The right panel compares the size of a proposed MOSAIC survey \citep{2024Msngr.193...24M,Gagnon-Hartman_2025} to pointings of the SKA-low and CCAT-prime.   From this figure we see that some obvious targets for cross correlations, such as the Subaru fields and COSMOS fields, are close to zero degrees.  This highlights the importance of being able to observe with SKA-low even at low latitudes.  Preliminary SKA observations must insure an overlapping footprint with such complimentary surveys.

\textbf{Author contributions}
A. Mesinger coordinated the writing of this chapter, edited the scientific content, and was responsible for the introduction. P. Dayal coordinated writing Sec. 2.1, and contributed to the science in other sections.
S.K. Giri coordinated writing Sec. \ref{sec:state_IGM}. T. R. Choudhury coordinated writing Sec. \ref{sec:4-1} and contributed to several other sections. B. Maity contributed to writing Sections 2.1, 2.2, and 2.3. Y. Qin coordinated writing Sec 2.3, and contributed to the scientific content of the other sections. A. Chakraborty contributed to the text in Sections 2.1 and 2.3.
C. Heneka coordinated and led the writing of section \ref{sec:3-2}.  S. Gagnon-Hartman, K. Moriwaki, and S. Yoshuira provided commenting and feedback. A. Gorce coordinated and led the writing of section \ref{sec:3_1}.
C. S. Murmu, S. Majumdar, and C. Heneka wrote most of the text in Section 3.3 of the chapter. A. Hutter coordinated and wrote most of Sec. \ref{sec:4-1} and contributed to Sec. \ref{sec:3-2}. K. Datta provided input to Sec. \ref{sec:4-1}. E. Zackrisson provided input and contributed to writing Sec. \ref{sec:4-1}. S. Gagnon-Hartman made Figure 10.  The other co-authors and the chapter lead further modified it to fit it seamlessly within the chapter. 

\bibliographystyle{abbrvnat-maxbibnames4}
\bibliography{chapter} % if your bibtex file is called example.bib

\end{document}